\documentclass[a4paper,11pt]{article}
\usepackage{amsmath,bm,amssymb,color}

\usepackage{graphicx}
\usepackage{jheppub} 
\usepackage{xspace}

\def\Q2{\left(Q^{2}\right)}

\def\d{{\rm d}}

\def\l({\left(}
\def\r){\right)}

\def\nf{N_{F}}
\def\CA{C_A}

\def\NF{N_F}

\def\d{\hbox{d}}

\def\EFTplus{EFT$\oplus$M}
\def\EFTtimes{EFT$\otimes$M}

\def\gapprox{\lower .7ex\hbox{$\;\stackrel{\textstyle >}{\sim}\;$}}
\def\lapprox{\lower .7ex\hbox{$\;\stackrel{\textstyle <}{\sim}\;$}}

\newcommand{\NNLOJET}{NNLO\protect\scalebox{0.8}{JET}\xspace}
\allowdisplaybreaks

\title{NNLO QCD corrections to Higgs boson production at large transverse momentum}

\author{X.\ Chen$^{a}$, J.\ Cruz-Martinez$^b$, T.\ Gehrmann$^{c}$, E.W.N.\ Glover$^b$, M.\ Jaquier$^d$}

\affiliation{
$^a$Center for High Energy Physics, Peking University, Beijing 100871, China\\
$^b$Institute for Particle Physics Phenomenology, Department of Physics, University of Durham, Durham, DH1 3LE, UK\\
$^c$Department of Physics, University of Z\"urich, CH-8057 Z\"urich, Switzerland\\
$^d$Albert-Ludwigs-Universit\"at Freiburg, Physikalisches Institut, D-79104 Freiburg, Germany}

\emailAdd{xuan.chen@pku.edu.cn}
\emailAdd{j.m.cruz-martinez@durham.ac.uk}
\emailAdd{thomas.gehrmann@uzh.ch}
\emailAdd{e.w.n.glover@durham.ac.uk}
\emailAdd{matthieu.jaquier@physik.uni-freiburg.de}

\abstract {We derive the second-order QCD corrections to the production of a Higgs boson recoiling against a parton with finite transverse momentum,
working in the effective field theory in which the top quark contributions are integrated out. To account for quark mass effects, we supplement the effective field theory result by the full quark mass dependence at leading order.
Our calculation is fully differential in the final state kinematics and includes the decay of the Higgs boson to a photon pair. It allows one to make next-to-next-to-leading order (NNLO)-accurate theory predictions for Higgs-plus-jet final states and for the transverse momentum distribution of the Higgs boson, 
accounting for the experimental definition of the fiducial cross sections. The NNLO QCD corrections are found to be moderate and positive, they 
lead to a substantial reduction of the theory uncertainty on the predictions.  We compare our results to 8 TeV LHC data from ATLAS and CMS. 
While the shape of the data is well-described for both experiments, we agree on the normalization only for CMS. By normalizing data and theory 
to the inclusive fiducial cross section for Higgs production, good agreement is found for both experiments, however at the expense of an increased theory uncertainty.  We make predictions for Higgs production observables at the $13$~TeV LHC, which are in good agreement with 
recent ATLAS data. At this energy,  the 
leading order 
mass corrections to the effective field theory prediction become significant at large transverse momenta, and we discuss the resulting uncertainties 
on the predictions.  }

\arxivnumber{1607.08817}

\begin{document}
\maketitle
\flushbottom

\section{Introduction}

Following the discovery of the Higgs boson~\cite{higgsexp} in 2012, the LHC experiments have now embarked on precision measurements 
of the Higgs boson properties, carried out by studying multiple production processes and decay modes. With more and 
more statistics collected by the experiments, differential
measurements in kinematical variables will become increasingly precise, thereby allowing detailed tests of the underlying Standard Model theory. 
A first glimpse at the potential of these studies can already be gained from the LHC 8 TeV data, with ATLAS~\cite{atlashpt} and 
CMS~\cite{cmshpt} performing, among other observables, 
 a first measurement of  the transverse momentum distribution of the Higgs boson in the diphoton decay mode. An important discriminator between 
Higgs production modes is the production of hadronic jets in association with the Higgs boson, with first cross section measurements 
available~\cite{atlashpt,cmshpt} from the 8 TeV data set. All these measurements are performed over a  fiducial region for the final state 
phase space of both the Higgs decay products and the final state jets. These fiducial cross sections are then essential ingredients for the extraction of 
total cross sections that are more easily converted into bounds on new physics effects through the determination of effective Higgs couplings. 
A precise theoretical description of fiducial cross sections in Higgs production, differential in kinematical variables and 
jet activity is therefore crucial for many upcoming studies in precision Higgs physics at the LHC. 

The dominant production mode of Higgs bosons at the CERN LHC is gluon fusion~\cite{ggH}, 
which is mediated through a heavy top quark loop. At leading order in perturbation theory,
${\cal O}(\alpha_s^2)$, the Higgs boson is always produced at zero transverse momentum. Perturbative higher order corrections to gluon fusion turn out to be numerically 
large and have been computed up to next-to-leading order (NLO) for full top quark mass dependence~\cite{spira,mtfinite} and to next-to-next-to-next-to-leading order (N3LO) in the limit of infinite top quark 
mass~\cite{ggHnlo,ggHnnlo,hnnlo,ggHn3lo}. Starting from NLO, the Higgs boson in gluon fusion can be produced recoiling against other final state partons, resulting in a 
finite transverse momentum of the Higgs boson. Consequently, the leading order (LO) process for Higgs production at non-vanishing transverse momentum is at 
${\cal O}(\alpha_s^3)$, and the 
counting of perturbative orders differs between inclusive Higgs production and the transverse 
momentum distribution of the Higgs boson. Including the full top quark mass dependence, the transverse momentum distribution is known only to LO~\cite{nigeljochum}, while 
NLO corrections were derived for infinite top quark mass~\cite{kunszt,ravindran}. 
The perturbative calculation of Higgs boson production at finite transverse momentum is closely related to Higgs-plus-jet production, and can be obtained from the latter 
by replacing  the kinematical requirements on the final state jet by an inclusive requirement on the total momentum of the final state partons, which counter-balance 
the Higgs boson. In the limit of infinite top quark mass, Higgs-plus-jet production was computed recently to next-to-next-to-leading order (NNLO), ${\cal O}(\alpha_s^5)$, by several 
groups~\cite{caolaH,ourH,caolaH2,njH,caolaH3}, using three 
different calculational approaches.  The NLO and NNLO predictions for the inclusive transverse momentum distribution has been further supplemented by the resummation of large logarithmic effects~\cite{MRT} up to next-to-next-to-leading level.

Together with the kinematical distributions, the experiments usually measure the total Higgs production cross section for the same 
 fiducial cuts on the Higgs boson decay products. By normalising the distributions to the total fiducial cross section, some of the experimental 
 uncertainties (mainly luminosity, but partly also reconstruction efficiencies and background subtractions) can be cancelled.

In this paper, we document our calculation of the NNLO corrections to Higgs-plus-jet production and extend it to describe the transverse momentum distribution 
of the Higgs boson to this order. These NNLO corrections are valid in the 
limit in which the top quark contributions are integrated out.  It is well known that finite quark mass effects are small when the partonic centre of mass energy is smaller than the top quark mass, and in particular when $p_T^H \lapprox m_t$.  However, at higher energies, and particularly at 
those that will be probed in Run 2 of the LHC, finite top mass effects are significant.     We therefore discuss how to supplement the NNLO effective field theory result by taking the full quark mass dependence into account at leading order yielding a more reliable prediction of the Higgs and jet transverse momentum distributions for $p_T^H \gapprox m_t$.

We focus on the relevant fiducial cross sections in the two-photon decay mode of the Higgs boson, 
which can be compared directly to experimental measurements without the need for 
an interpolation of data into unmeasured regions. The paper is structured as follows: in section~\ref{sec:setup}, we describe the setup of the calculation, which is first 
validated in Higgs-plus-jet final states in section~\ref{sec:hjet} where we also compare to the Run 1 data from 
ATLAS~\cite{atlashpt} and CMS~\cite{cmshpt}, and discuss the 
impact of considering normalised cross sections.  Section~\ref{sec:hpt} contains a detailed discussion of the NNLO corrections to Higgs boson production 
cross sections at moderate transverse momentum, and a comparison with existing Run 1 data from the ATLAS~\cite{atlashpt} and CMS~\cite{cmshpt} experiments. First preliminary ATLAS results~\cite{atlasICHEP} 
from Run 2 at 13 TeV are discussed in section~\ref{sec:ichep}. 
We then turn in section~\ref{sec:hpt2} to the production of Higgs bosons at the larger transverse momenta that will be probed in Run 2 of the 13~TeV LHC where the top quark mass effects are substantial. 
We conclude with an outlook in Section~\ref{sec:conc}.

\section{Setup of the calculation}
\label{sec:setup}

Higgs production in gluon fusion is mediated through a heavy top quark loop. If all scales involved in the process under consideration are substantially smaller than 
the top quark mass, it is possible to integrate out the top quark loop by taking the limit $m_t\to \infty$. The resulting effective field theory (EFT) 
Lagrangian~\cite{eft} then consists of five-flavour QCD and a term coupling the Higgs field to the square of the gluon field strength tensor,
\begin{equation}
{\cal L}_{EFT} = -\frac{\lambda}{4} G^{\mu\nu}G_{\mu\nu} H.
\end{equation}
The matching of this EFT onto full QCD and its renormalisation have been derived to three-loop order~\cite{steinhauser}. The presence of the effective 
coupling $\lambda$ alters the renormalization scale dependence of the hard subprocess cross section, compared to full QCD. 

For a fixed renormalization scale 
$\mu_0$, the cross section for single Higgs production processes via gluon fusion with $n$ final state jets at leading order 
can be written in terms of the effective coupling 
$\lambda$ as:
\begin{eqnarray}
\label{eq:higgsxs}
\lefteqn{\sigma^{H+nJ}(\mu_0,\alpha_s(\mu_0),\lambda(\mu_0)) 
= \lambda(\mu_0)^2 }\nonumber \\
&\times
\left[
\left(\frac{\alpha_s(\mu_0)}{2\pi}\right)^{n-2}\sigma^{(0)} 
+\left(\frac{\alpha_s(\mu_0)}{2\pi}\right)^{n-1}  \sigma^{(1)}(\mu_0)
+\left(\frac{\alpha_s(\mu_0)}{2\pi}\right)^{n} \sigma^{(2)}(\mu_0) 
+{\cal O} (\alpha_s^{n+3}) \right]
\end{eqnarray}
where,
\begin{equation}
	\lambda(\mu_0)^2 = C_0 \left(\frac{\alpha_s(\mu_0)}{2\pi}\right)^{2}
\left(1 
+ C_1(\mu) \left(\frac{\alpha_s(\mu_0)}{2\pi}\right) 
+ C_2(\mu) \left(\frac{\alpha_s(\mu_0)}{2\pi}\right)^2 
\right)\label{eq:effCoup}
\end{equation}
with,
\begin{eqnarray}
\label{eq:C0}   C_0 &=& \frac{4}{9v^2},\\
\label{eq:C1} C_1(\mu_0) &=& \frac{11N}{3} \equiv C_1, \\
\label{eq:C2} C_2(\mu_0) &=& \frac{1933N^2}{162} - \frac{67}{108}\nf + (C_1\beta_0 - 2\beta_1)
\log\left(\frac{\mu_0^2}{m_t^2}\right) \,,
\end{eqnarray}
and the one- and two-loop QCD beta functions $\beta_0$ and $\beta_1$ are given in Appendix~\ref{app:scales} in Eq.~\eqref{eq:betas}.
Expansion of the effective coupling in $\alpha_s$ in~\eqref{eq:higgsxs} yields
\begin{eqnarray}
\label{eq:higgsxs2}
\lefteqn{
\sigma^{H+nJ}(\mu_0) 
=\phantom{+} C_0 \left(\frac{\alpha_s(\mu_0)}{2\pi}\right)^{n} \sigma^{(0)}
 + C_0  \left(\frac{\alpha_s(\mu_0)}{2\pi}\right)^{n+1} \left(\sigma^{(1)}(\mu_0) + C_1 \sigma^{(0)}\right)
}\nonumber \\
&& + C_0  \left(\frac{\alpha_s(\mu_0)}{2\pi}\right)^{n+2} \left( \sigma^{(2)}(\mu_0) + C_1 \sigma^{(1)}(\mu_0)+C_2(\mu_0) \sigma^{(0)}\right)
+{\cal O} (\alpha_s^{n+3})\,.
\end{eqnarray}

The scale dependence of the cross section \eqref{eq:higgsxs} can then be reconstructed from the running of $\alpha_s$, 
\begin{eqnarray}
\lefteqn{
\sigma^{H+nJ}(\mu_R) 
\label{eq:higgsxs4}
=\phantom{+} C_0 \left(\frac{\alpha_s(\mu_R)}{2\pi}\right)^{n} \sigma^{(0)}
+ C_0  \left(\frac{\alpha_s(\mu_R)}{2\pi}\right)^{n+1} \left(
\sigma^{(1)}(\mu_R) + C_1 \sigma^{(0)} \right)}\nonumber \\
&& + C_0  \left(\frac{\alpha_s(\mu_R)}{2\pi}\right)^{n+2} \left( \sigma^{(2)}(\mu_R) + C_1 \sigma^{(1)}(\mu_R)+C_2(\mu_R) \sigma^{(0)} \right)
+{\cal O} (\alpha_s^{n+3}) \,
\end{eqnarray}
where we have identified,
\begin{eqnarray}
	\sigma^{(1)}(\mu_R) & =& \sigma^{(1)}(\mu_0) + n\beta_0L_R\sigma^{(0)}\label{eq:si1scales}\\
	C_2(\mu_R)       & = & C_2(\mu_0) + L_R\beta_0 C_1 - 2\beta_1L_R\label{eq:wc2r}
\end{eqnarray}
and crucially,
\begin{eqnarray}
\sigma^{(2)}(\mu_R) &=& \sigma^{(2)}(\mu_0) + (n+1)\beta_0L_R\sigma^{(1)}(\mu_0)  \nonumber \\
					 &&+ \left((n+2)\beta_1 L_R + \frac{n(n+1)}{2}\beta_0^2L_R^2\right)\sigma^{(0)} \label{eq:si2scales}
\end{eqnarray}
Comparing this  to the scale dependence obtained in full QCD~(see Eq.~\eqref{eq:mRvar}), we observe that the 
 scale dependence of $\sigma^{(2)}(\mu_R)$ has been altered to absorb the $\beta_1$ contribution in $C_2(\mu_R)$.
The LO, NLO and NNLO expressions for the cross sections are obtained by truncating (\ref{eq:higgsxs4}) to the respective orders in 
$\alpha_s(\mu_R)$. 

\begin{table}[b]
\begin{tabular}{lll}
\hline
LO & $gg \to Hg$, $qg \to Hq$, $q\bar q \to Hg$ & tree level\\ \hline 
NLO & $gg \to Hg$, $qg \to Hq$, $q\bar q \to Hg$ & one loop\\
& $gg \to Hgg$, $gg\to Hq\bar q$, $qg \to Hqg$, & tree level\\
&  $qq \to H qq$, 
 $q\bar q \to Hgg$, $q\bar q \to H q\bar q$ &  \\ \hline
 NNLO & $gg \to Hg$, $qg \to Hq$, $q\bar q \to Hg$ & two loop\\
& $gg \to Hgg$, $gg\to Hq\bar q$, $qg \to Hqg$, & one loop\\
&  $qq \to H qq$, 
 $q\bar q \to Hgg$, $q\bar q \to H q\bar q$ &   \\
&  $gg \to Hggg$, $gg\to Hq\bar qg$, $qg \to Hqgg$, & tree level \\
& $qg \to H qq\bar q$, $q q \to Hqqg$, $q\bar q \to H ggg$, &\\
 & $q\bar q \to H q\bar q g$\\ \hline 
\end{tabular}
\caption{Parton-level processes contributing to Higgs boson production at finite transverse momentum in different orders in perturbation theory.~\label{tab:processes}}
\end{table}
Our calculation of the NNLO corrections to Higgs boson production at finite transverse momentum is performed within the EFT framework. 
 The process receives leading-order contributions from the parton level processes $gg\to Hg$, $qg \to Hq$ and $q\bar q\to Hg$, where 
the two former account for the bulk of the cross section. The contributions at higher orders are summarized in Table~\ref{tab:processes}, the relevant matrix elements at 
tree level~\cite{h5g0l}, one loop~\cite{h4g1l} and two loops~\cite{h3g2l} were computed already a while ago. 
The ultraviolet renormalised matrix elements for these processes are integrated  over the final state phase space appropriate to 
Higgs boson final states, including a 
cut on either $p_{T}^{H}$ or $p_T^j$. All three types of contributions are infrared-divergent and only their sum is finite.  

In this calculation we employ the antenna subtraction method~\cite{ourant} to isolate the infrared singularities 
in the different Higgs-boson-plus-jet contributions to enable their cancellation prior to the numerical implementation. 
The construction of the subtraction terms is exactly as described in Ref.~\cite{ourH} for the gluons-only subprocess, 
now including all 
 partonic channels relevant to Higgs-boson-plus-jet production. Our calculation is 
implemented in a newly developed parton-level Monte Carlo generator \NNLOJET. This program
provides the necessary 
infrastructure for the antenna subtraction of hadron collider processes at NNLO and performs the integration 
of all contributing subprocesses at this order. Components of it have also been used in 
other NNLO QCD calculations~\cite{eerad3,nnlo2j,nnlott,nnlozj,nnlodis} using the antenna subtraction method. 
Other processes can be added to \NNLOJET provided the matrix elements are 
available. 

To describe the normalised distributions, we also implemented 
the NNLO QCD corrections to inclusive Higgs boson production including the decay to photon pairs in \NNLOJET
and validated this 
implementation for fiducial cross sections against the publicly available HNNLO code~\cite{hnnlo}.

For our numerical computations, we take the  Higgs boson mass  $m_H= 125$~GeV and the 
vacuum expectation value  $v=246.2$~GeV. To estimate massive quark effects, 
the three heaviest quarks are considered with masses: {$m_t = 173.2$~GeV, $m_b =4.18$~GeV, $m_c =1.275$~GeV.}
We use the PDF4LHC15 parton distribution functions (PDFs)~\cite{nnpdf}
with the value of $\alpha_s(M_Z)=0.118$ at NNLO, and $M_Z=91.1876~$GeV. Note that we systematically use the 
same family of PDFs and the same value of $\alpha_s(M_Z)$ for the NLO (PDF4LHC15\_nlo\_30) and NNLO (PDF4LHC15\_nnlo\_30) predictions. The factorisation and renormalisation scales are chosen dynamically on an event-by-event basis as,
\begin{equation}
  \label{eq:scale}
\mu \equiv \mu_R = \mu_F = \frac{1}{2}\sqrt{m_H^2 + (p^H_T)^2},
\end{equation}
where $m_H$ and $p^H_T$ are the invariant mass and the transverse momentum of the final state photon pair respectively. The theoretical uncertainty is estimated by varying the scale choice by a factor in the range $[1/2,2]$.

The large transverse momentum region is of fundamental interest in view of possible deviations from Standard Model expectations~\cite{bsm}. New physics effects 
can modify the transverse momentum distribution of the Higgs boson either directly through new Higgs production processes in the decay of new heavy 
particles, or indirectly through the presence of new massive states in the loop that couples the Higgs boson to gluons. At high transverse momentum,
this heavy particle loop is resolved by the large momentum transfer flowing through it, and the EFT description that reduces the loop to a 
point-like coupling is no longer applicable. For the Standard Model prediction, this implies that the dependence of the cross section on the 
top quark mass can no longer be neglected at large transverse momenta, $p_T^H \sim m_t$. At present, exact expressions for the matrix elements 
for Higgs production at finite transverse momentum are known only at one-loop~\cite{nigeljochum}, which amounts to the LO contribution. 
Higher-order terms in a mass expansion were computed in~\cite{harlander}, where re-weighting procedures for the NLO EFT
for Higgs-plus-jet production and the Higgs transverse momentum distribution are compared in detail. 

For a fully consistent description of the mass effects at high transverse momentum, one would like to have the NLO (and ultimately also NNLO)
predictions with exact mass dependence. However, owing to the complexity of the two-loop virtual amplitudes, these are not available at present. We therefore introduce two approximate approaches to estimating the mass effects.
For the inclusive Higgs production cross section, it has been observed that the top quark mass corrections at NLO~\cite{spira,mtfinite} can be 
well-approximated  by re-weighting the NLO EFT cross section by the ratio of full and EFT predictions at leading order:
\begin{equation}
\label{eq:Rdef}
R = \sigma_{{\rm LO}}^M / \sigma_{{\rm LO}}^{{\rm EFT}}
\end{equation}
where $\sigma_{{\rm LO}}^M$ includes the exact mass dependence of top quark loops. The numerically smaller 
contributions from charm and bottom quarks to 
the inclusive Higgs boson cross section can be accounted for in the same form by including the relevant quark loops in the numerator.
In the following, the re-weighting factor $R$ will always include charm, bottom and top loops in $\sigma_{{\rm LO}}^M$, while normalising to 
$\sigma_{{\rm LO}}^{{\rm EFT}}$ with infinite top quark mass. In the EFT, all other quarks are treated massless, and their Yukawa couplings are set to zero.

This inclusive re-weighting factor can be generalised to the transverse momentum distribution (which is also inclusive in all hadronic radiation) as
\begin{equation}
R(p^H_T) = \left(\frac{\d \sigma_{{\rm LO}}^M}{\d p^H_T}\right) \Bigg/ \left(\frac{\d \sigma_{{\rm LO}}^{{\rm EFT}}}{\d p^H_T}\right).
\label{eq:rmass}
\end{equation}
Multiplying the higher order EFT predictions bin-by-bin with this factor, yields the \EFTtimes\ approximation
\begin{equation}
\label{eq:Mtimes}
\frac{\d \sigma_{{\rm NNLO}}^{EFT\otimes M}}{\d p^H_T} \equiv R(p^H_T) \left(\frac{\d \sigma_{{\rm NNLO}}^{{\rm EFT}}}{\d p^H_T}\right)  
\end{equation}
which correctly captures the leading logarithms in the quark mass corrections~\cite{ptmass} at all orders, while failing in general to describe subleading 
logarithms and non-logarithmic terms. The computation of subleading mass corrections at NLO~\cite{harlander,Dawson:2014ora} also suggests the applicability of 
the \EFTtimes\ procedure. 
To quantify the uncertainty associated with  this re-weighting procedure, we consider also
the additive \EFTplus\ prediction obtained by substituting only the LO EFT contribution by the full LO mass-dependence,
\begin{equation}
\label{eq:Mplus}
\frac{\d \sigma_{{\rm NNLO}}^{EFT\oplus M}}{\d p^H_T} \equiv  \left(\frac{\d \sigma_{{\rm NNLO}}^{{\rm EFT}}}{\d p^H_T}\right) + \left (R(p^H_T) -1\right) \left(\frac{\d \sigma_{{\rm LO}}^{{\rm EFT}}}{\d p^H_T}\right).
\end{equation}

\begin{figure}
  \centering
\includegraphics[width=7cm,natwidth=610,natheight=642]{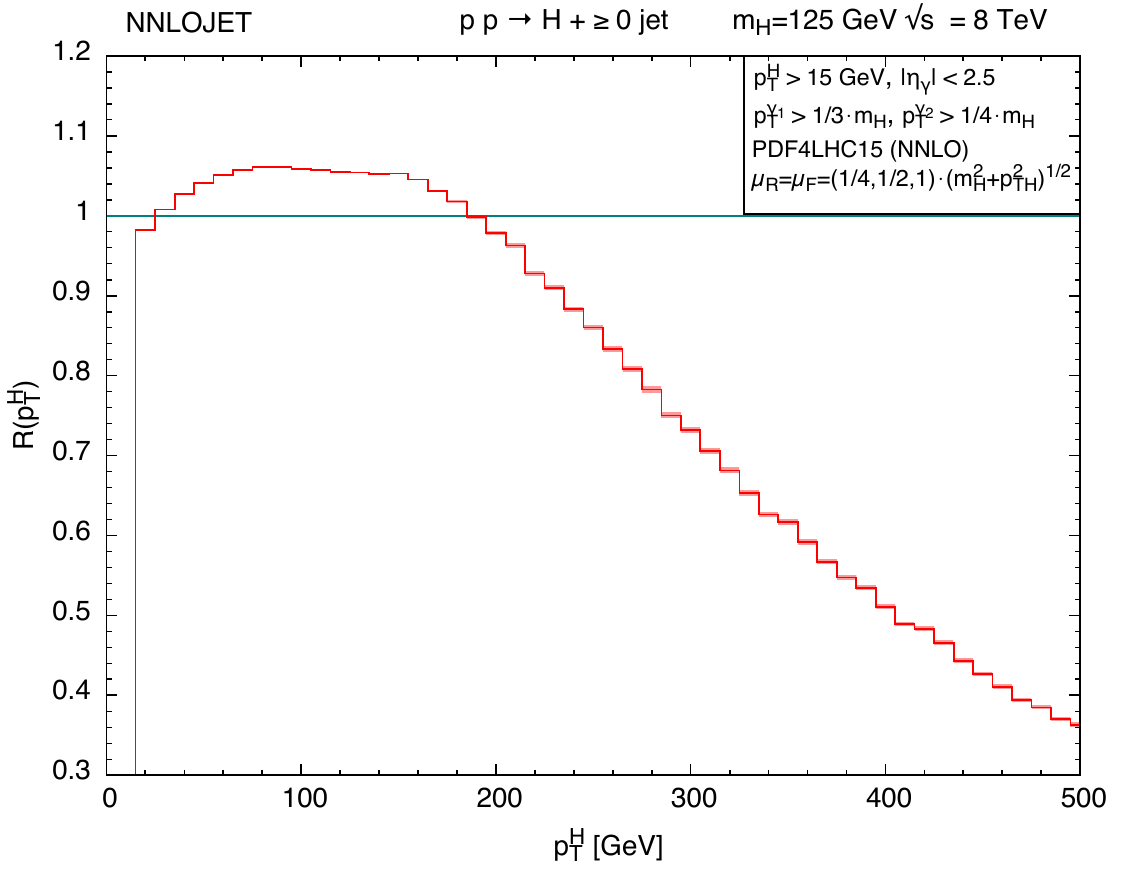}
\includegraphics[width=7cm,natwidth=610,natheight=642]{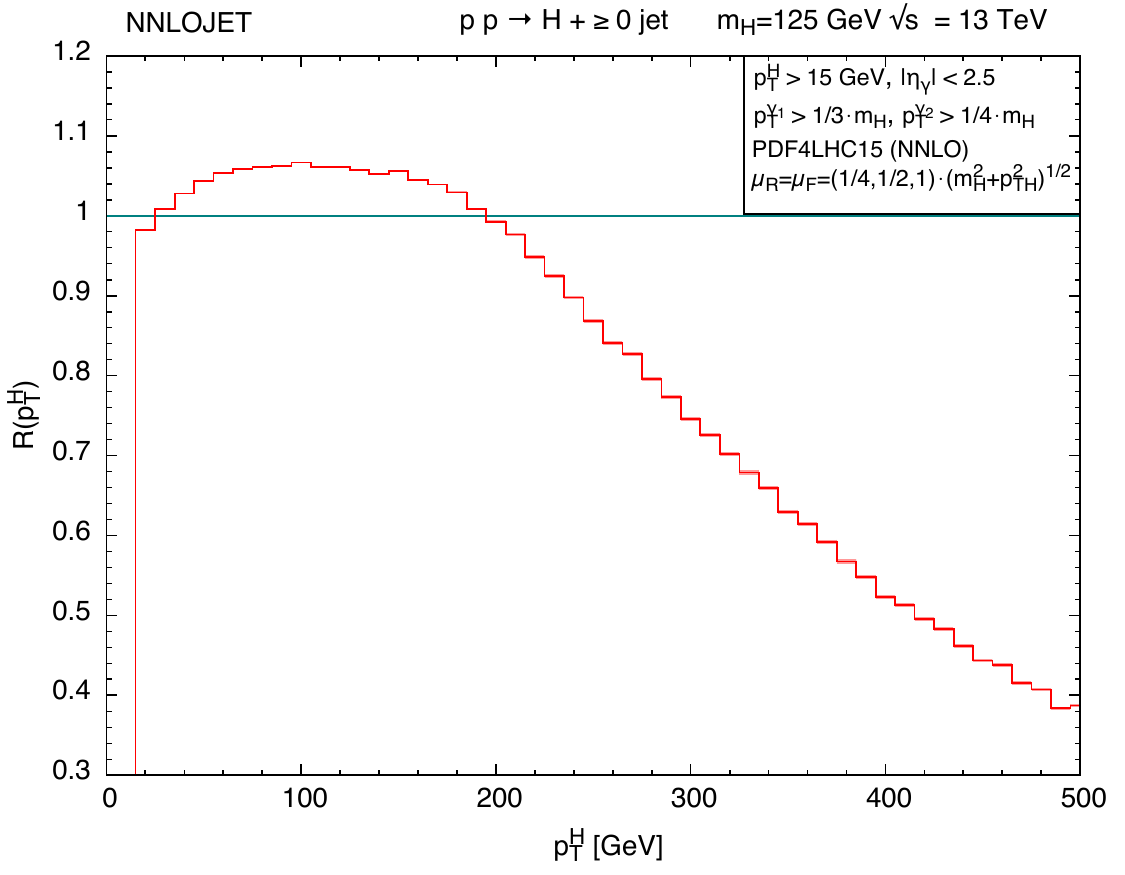}
\caption{The scaling factor $R(p^H_T)$ (for a representative set of fiducial cuts) at leading order with exact quark mass 
dependence at 8 TeV (left) and at 13 TeV (right). 
\label{fig:masslo8and13}}
\end{figure}

To quantify the impact of the top, bottom and charm quark mass effects we consider 
a representative set of fiducial cuts applied at 8 and 13 TeV. Figure~\ref{fig:masslo8and13} shows $R(p^H_T)$ as a function of $p^H_T$.
We observe that the exact quark mass dependence leads to a mild enhancement (about 4.5\% at 8 TeV and about 5.5\% at 13 TeV) in the 
transverse momentum range up to about $m_t$. Above $p^H_T \sim m_t$, $R(p^H_T)$ falls off steeply with increasing 
transverse momentum as the top quark circulating in the loops starts to become resolved.
The inclusive reweighting factor, defined according to Eq.~\eqref{eq:Rdef}, is $R = 0.936$.
 It depends only on the ratio of the quark masses to the Higgs boson mass, 
 and is thus independent of the collider energy and the fiducial cuts on the photons.

Figure~\ref{fig:masslo8and13} shows that the mass effects are generally positive and amount to a few percent in the kinematic regions probed by ATLAS and CMS in Run 1, leading to a small enhancement of the cross section.  Owing to the smallness of the effect, the \EFTplus\ and \EFTtimes\ lead to very similar predictions.  Therefore, at 8 TeV, we systematically make theoretical predictions using the \EFTtimes\ approximation. 

However, for the kinematic regions one expects to probe in Run 2 of the LHC at 13 TeV, the massive quark effects become much more important and we will systematically study the uncertainty that the mass effects introduce in the NNLO predictions in the EFT, the \EFTtimes\ and \EFTplus\ approximations.  

\section{Higgs production at moderate transverse momentum}

Higgs production at moderate transverse momentum 
has been studied by ATLAS~\cite{atlashpt} and CMS~\cite{cmshpt}, based on the data taken at 8 TeV, especially in the  diphoton decay mode of the Higgs boson, which allows a full kinematical reconstruction. The measurements are performed in fiducial phase space regions, to ensure that all final state objects (jets, photons) are well within the detector 
coverage and can be reconstructed reliably. The resulting data provide the first-ever measurement of Higgs boson production at moderate transverse momentum including
Higgs-plus-jet final states; they demonstrate the future potential of this type of observables and allow for detailed comparisons between data and theory. 
\begin{table}
\begin{center}
\begin{tabular}{ccc}
& ATLAS & CMS\\ \hline
leading photon  & $|\eta_{\gamma_1}|<2.37$ & $|\eta_{\gamma_1}|<2.5$  \\
& $p^{\gamma_1}_{T} > 0.35\,m_H$ &  $p^{\gamma_1}_{T} > 0.33\,m_H$ \\
 sub-leading photon & $|\eta_{\gamma_2}|<2.37$ &  $|\eta_{\gamma_2}|<2.5$  \\
 & $p^{\gamma_2}_{T} > 0.25\,m_H$  &   $p^{\gamma_2}_{T} > 0.25\,m_H$  \\
photon isolation & $R_{\gamma}=0.4$ & $R_{\gamma}=0.4$\\
& $\sum_i E_{Ti} < 14$~GeV & $\sum_i E_{Ti} < 10$~GeV\\
anti-$k_T$ jets & $R=0.4$ & $R=0.5$\\
& $|\eta_j| < 4.4$  & $|\eta_j| < 2.5$   \\
 & $p^{j}_{T} > 30$~GeV &  $p^{j}_{T} > 25$~GeV \\
\hline
\end{tabular}
\end{center}
\caption{Kinematical cuts used to define the fiducial phase space for the final state photons and jets in the measurements of ATLAS~\protect{\cite{atlashpt}} and CMS~\protect{\cite{cmshpt}}. The measurements of the total fiducial cross section and of the inclusive transverse momentum distribution do not apply the jet cuts. 
\label{tab:cuts}}
\end{table}

The fiducial event selection cuts for the ATLAS and CMS measurements of Higgs-plus-jet production in the diphoton decay mode are summarized in 
Table~\ref{tab:cuts}. To mimic the photon isolation cuts, we limit the sum of partonic transverse energy deposited close to the photon, $\sum_i E_{Ti}$, where $i$ runs over all the final state partons within a distance $R_\gamma$ of the photon. 

Normalized distributions are obtained by dividing the experimental data by the measured total cross sections in the fiducial region for 
inclusive Higgs production $\sigma_H$ and by dividing the theory predictions by  $\sigma_H$
evaluated in the corresponding effective theory (EFT, \EFTtimes\ and \EFTplus) to NNLO accuracy (${\cal O}(\alpha_s^4)$), see Table~\ref{tab:xsec}. To estimate the theoretical scale uncertainty for normalized cross sections, we use the same scale choice as in Eq.~\eqref{eq:scale} and vary the scales in the range $[1/2,2]$ independently in the numerator 
and denominator. 
\begin{table}
\begin{center}
\begin{tabular}{ccc}
& ATLAS & CMS\\ \hline
$\sigma_{H,{\rm exp}}$ 
& $43.2 \pm 9.4 {+3.2 \atop -2.9} \pm 1.2$~fb & $32.2  {+10.1 \atop -9.7} \pm 3.0$~fb\\
$\sigma_{H,{\rm NNLO}}^{EFT}$ & $27.0 {+1.3 \atop -2}$~fb & $28.2 {+1.4 \atop -2.1}$~fb \\ 
$\sigma_{H,{\rm NNLO}}^{EFT\otimes M}$ & $25.2 {+1.2 \atop -1.9}$~fb & $26.4 {+1.3 \atop -1.9}$~fb \\ 
$\sigma_{H,{\rm NNLO}}^{EFT\oplus M}$ & $26.3 {+1.2 \atop -1.9}$~fb & $27.5 {+1.2 \atop -2}$~fb \\ \hline
$\sigma_{H+\ge 1 jet,{\rm exp}}$ 
& $21.5 \pm 5.3{+2.4 \atop -2.2} \pm 0.6$~fb &  - \\
$\sigma_{H+\ge 1 jet,{\rm NNLO}}^{EFT}$ & $9.5 {+0.03 \atop -0.69}$~fb & $10.3 {+0.21 \atop -0.85}$~fb \\ 
$\sigma_{H+\ge 1 jet,{\rm NNLO}}^{EFT\otimes M}$ & $9.8 {+0.04 \atop -0.71}$~fb & $10.6 {+0.22 \atop -0.88}$~fb \\ 
$\sigma_{H+\ge 1 jet,{\rm NNLO}}^{EFT \oplus M}$ & $9.7 {+0.11 \atop -0.73}$~fb & $10.5 {+0.27 \atop -0.89}$~fb \\ 
\hline
$\sigma_{H+ 1 jet,{\rm exp}}$ 
& $12.3{+4.7 \atop -4.8}$ ~fb & $4.3 {+6.4 \atop -6.3 }$~fb\\
$\sigma_{H+ 1 jet,{\rm NNLO}}^{EFT}$ & $6.8 {-0.10 \atop -0.19}$~fb & $7.5 {-0.21 \atop -0.25}$~fb \\ 
$\sigma_{H+ 1 jet,{\rm NNLO}}^{EFT\otimes M}$ & $7.0 {-0.10 \atop -0.18}$~fb & $7.7 {-0.21 \atop -0.26}$~fb \\ 
$\sigma_{H+ 1 jet,{\rm NNLO}}^{EFT\oplus M}$ & $7.0 {-0.02 \atop -0.22}$~fb & $7.6 {-0.15 \atop -0.29}$~fb \\ 
\hline
\end{tabular}
\end{center}
\caption{Fiducial inclusive cross sections used for the normalization of the distributions (upper). Fiducial inclusive cross sections for Higgs+jet (middle). Fiducial exclusive cross sections for Higgs+jet (lower). Experimental errors are statistical, systematical and luminosity (ATLAS only). Theoretical uncertainties for the EFT, \EFTtimes\ and \EFTplus\ approximations are from scale variation as described in the text.
\label{tab:xsec}}
\end{table}

\subsection{Higgs boson plus jet production}
\label{sec:hjet}

A substantial fraction of Higgs bosons produced in gluon fusion are accompanied by hadronic jets. 
The NNLO corrections to Higgs-plus-jet production were initially derived for the gluons-only subprocess using 
a sector-improved residue subtraction scheme~\cite{stripper} in Ref.~\cite{caolaH} and using 
antenna subtraction~\cite{ourant} in Ref.~\cite{ourH}. This subprocess alone is however insufficient for a full 
phenomenological description, it was later on extended to a full calculation~\cite{caolaH2} in 
sector-improved residue subtraction, and applied 
to compute the fiducial cross sections 
measured by ATLAS~\cite{atlashpt} in~\cite{caolaH2}. 
An independent calculation~\cite{njH} used the newly developed 
NJettiness subtraction method~\cite{njet}.  To validate our code, 
we made an in-depth comparison (using NNPDF2.3 PDFs for ATLAS cuts but without the photon isolation requirement) with the calculation of~\cite{caolaH2}. When properly accounting for the omission of the numerically small $q\bar q$ channel (NNLO only) in~\cite{caolaH2}, we found agreement to better than 
5 per mille on the NNLO cross sections. 
For the total Higgs-plus-jet cross section, we cross checked three different set of cuts and listed the details in table~\ref{tab:crosscheckxsec}. We found good agreement of the total Higgs-plus-jet cross sections with~\cite{caolaH3} and~\cite{hjveto}. We also attempted a comparison with the published numbers in~\cite{njH}, but were unable to confirm them. 

\begin{table}
\begin{center}
\resizebox{\textwidth}{!}{
\begin{tabular}{cccc}
 \hline
$\sqrt{s}$
& 8 TeV & 13 TeV & 8 TeV\\
PDF set
& NNPDF23$\_$nnlo & PDF4LHC15$\_$nnlo$\_$30 & NNPDF23$\_$nnlo\\
Central scales 
& $\mu_R=\mu_F=m_H$ & $\mu_R=\mu_F=m_H$ & $\mu_R=\mu_F=m_H$ \\ 
anti-$k_T$ jets 
& $R=0.4$ & $R=0.4$ & $R=0.5$ \\
& $|\eta_j| < 4.4$ & - & $|\eta_j| < 2.5$ \\ 
& $p^{j}_{T} > 30$~GeV & $p^{j}_{T} > 30$~GeV & $p^{j}_{T} > 30$~GeV\\
leading photon
& $|\eta_{\gamma_1}|<2.37$ & - & -  \\
& $p^{\gamma_1}_{T} > 0.35\,m_H$ & - & -  \\
sub-leading photon  
& $|\eta_{\gamma_2}|<2.37$ & - & -  \\
& $p^{\gamma_2}_{T} > 0.25\,m_H$ & - & -  \\
Parton channels
& $gg$+$qg$+$q\bar{q}$(NLO) & $gg$+$qg$+$q\bar{q}$(NLO) & all channels (NNLO)  \\
\hline
&$\sigma_{H(\rightarrow \gamma\gamma)+\ge 1 jet,{\rm NNLO}}^{EFT}$ & $\sigma_{H+\ge 1 jet,{\rm NNLO}}^{EFT}$ & $\sigma_{H+\ge 1 jet,{\rm NNLO}}^{EFT}$ \\
 NNLOJET
& $9.44^{+0.59}_{-0.85}$ fb & $16.8^{+0.9}_{-1.5}$ pb & $5.81^{+0.51}_{-0.62}$ pb  \\
Results from ~\cite{caolaH3}
& $9.45^{+0.58}_{-0.82}$ fb & - & -  \\
Results from ~\cite{hjveto}
& - & $16.7^{+1.0}_{- -}$ pb & -  \\
Results from ~\cite{njH}
& - & - & $5.5^{+0.3}_{-0.4}$ pb  \\
\hline
\end{tabular}
}
\end{center}
\caption{Comparison of NNLOJET results for Higgs-plus-jet cross sections at NNLO with previous 
results in the literature~\cite{caolaH3,hjveto,njH}, with fiducial cuts, parton distributions and parton-level channels as in the 
respective studies. The theoretical uncertainty is estimated by varying the central scale by a factor in the range [1/2, 2]. In~\cite{hjveto}, the 
cross section at $\mu_R=\mu_F=2m_H$ scale is not quoted. 
\label{tab:crosscheckxsec}}
\end{table}

\begin{figure}[t]
  \centering
\includegraphics[width=7cm,natwidth=610,natheight=642]{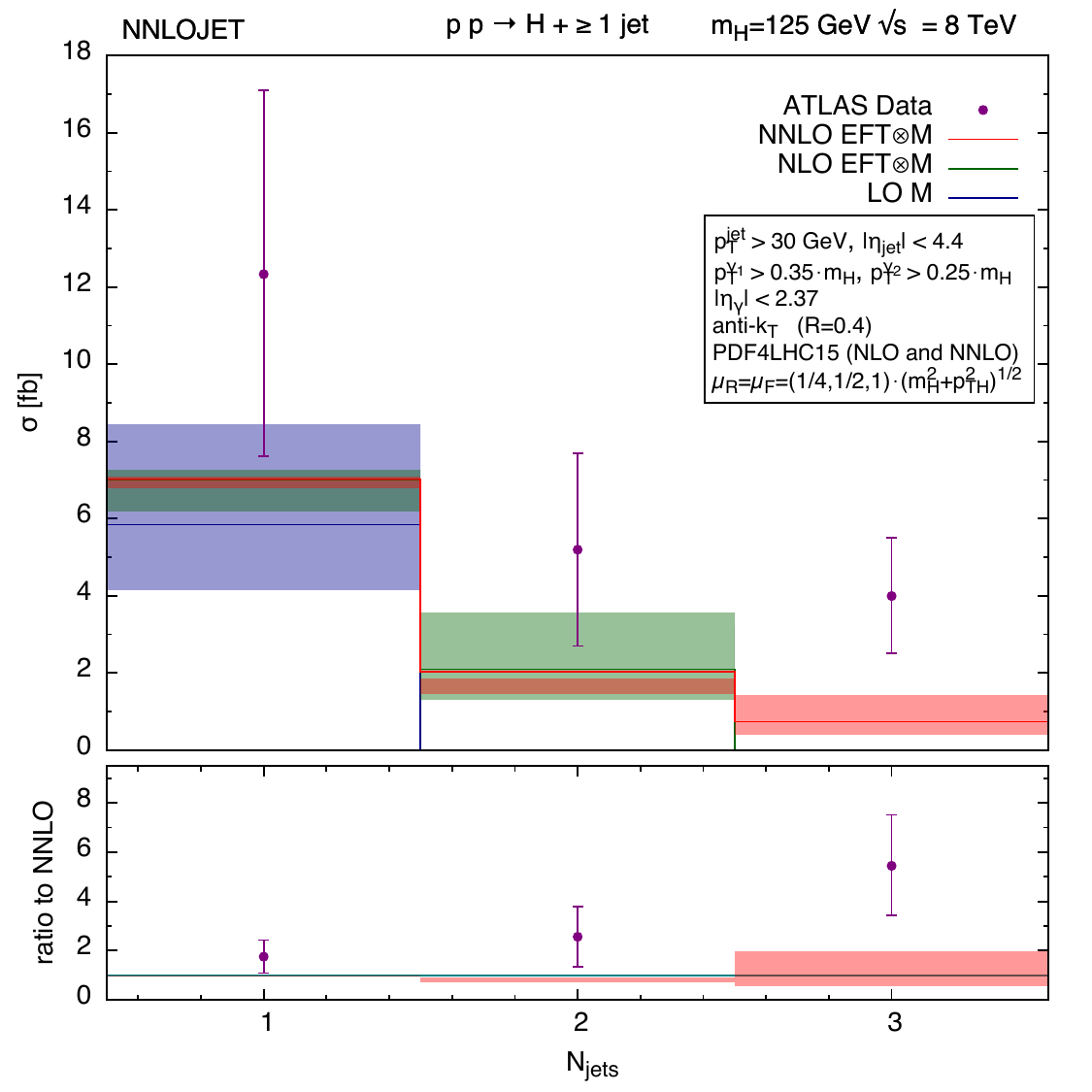}
\includegraphics[width=7cm,natwidth=610,natheight=642]{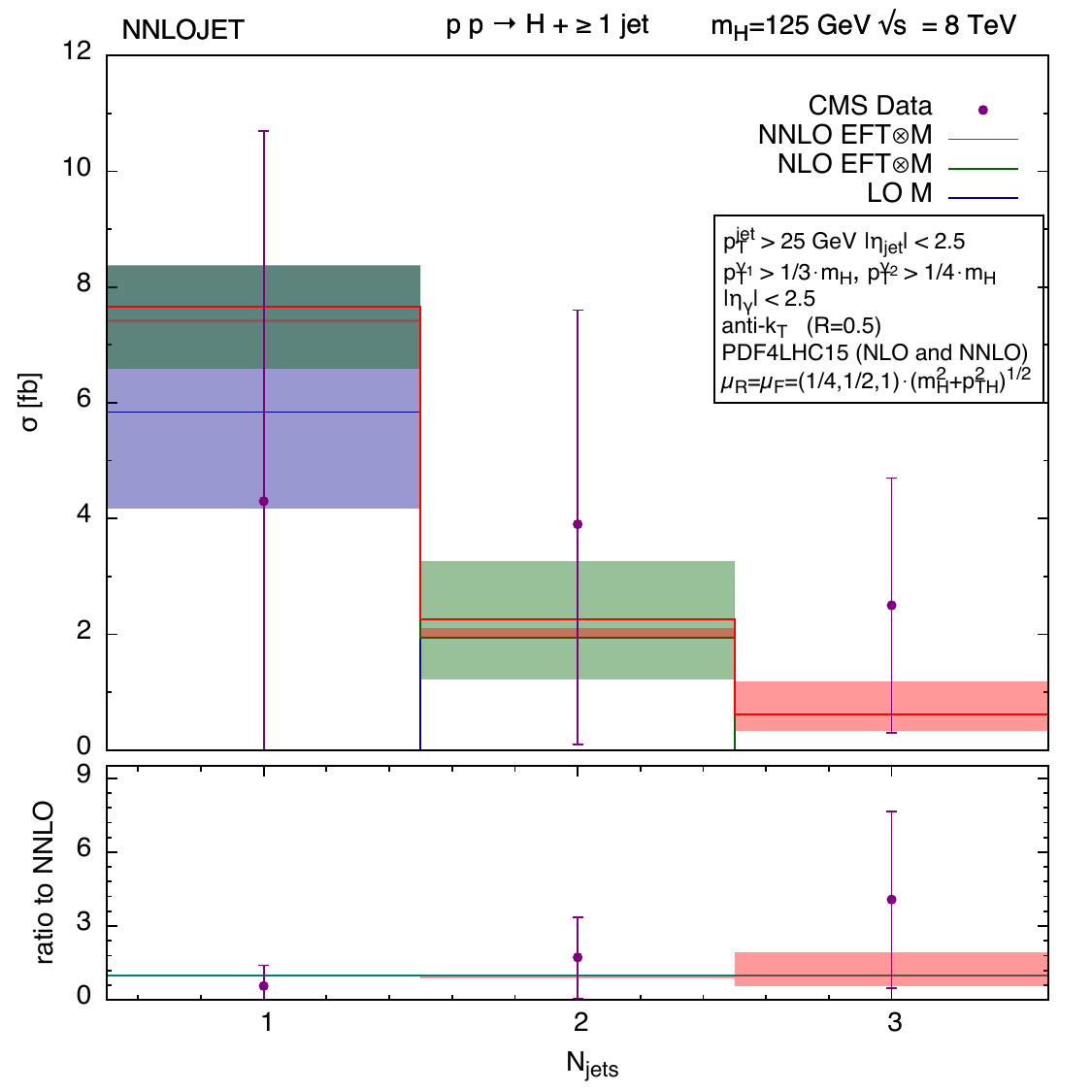}
\caption{Jet multiplicity in Higgs-plus-jet production compared to ATLAS~\protect{\cite{atlashpt}} and CMS~\protect{\cite{cmshpt}} data.\label{fig:njet}}
\vspace{4mm}
  \centering
\includegraphics[width=7cm,natwidth=610,natheight=642]{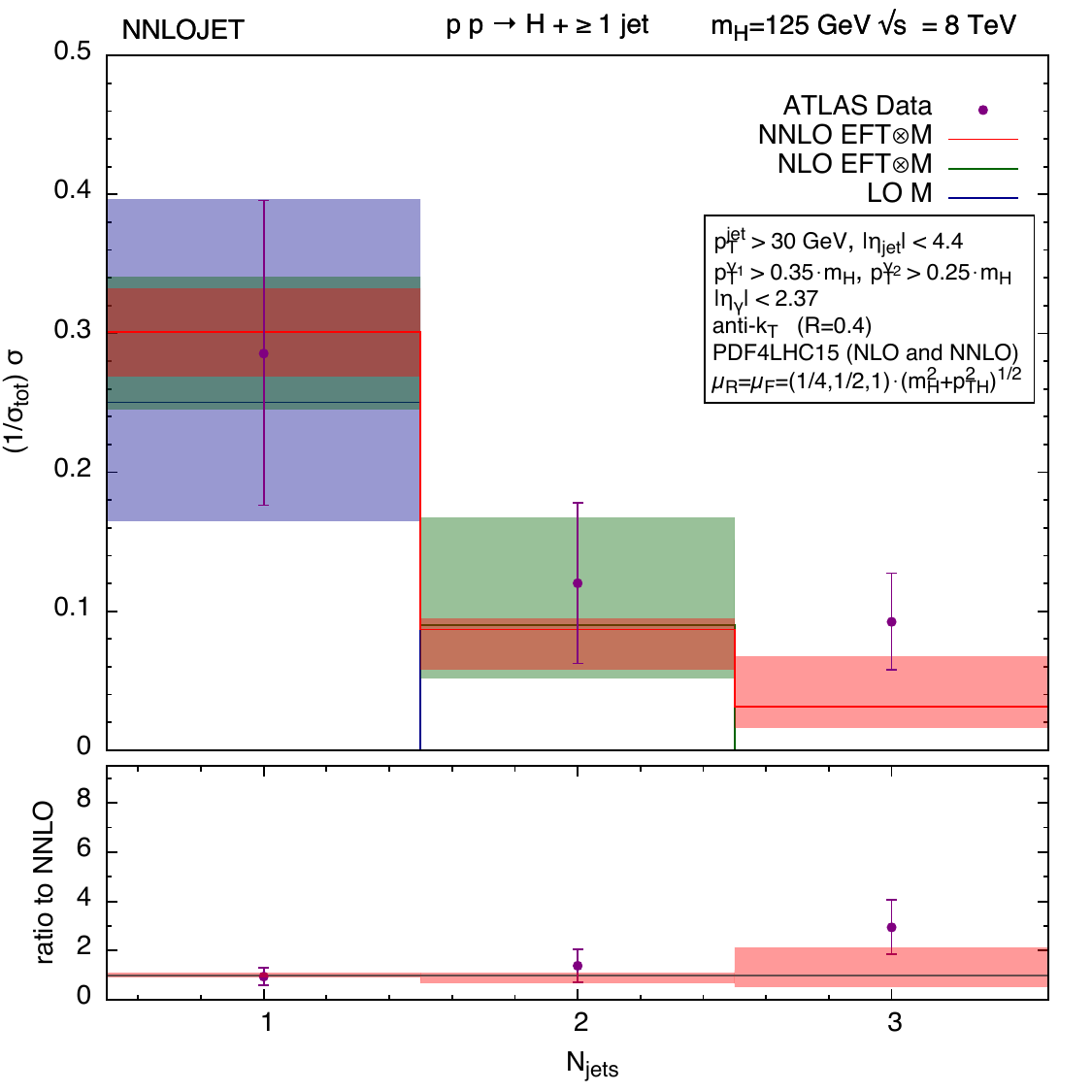}
\includegraphics[width=7cm,natwidth=610,natheight=642]{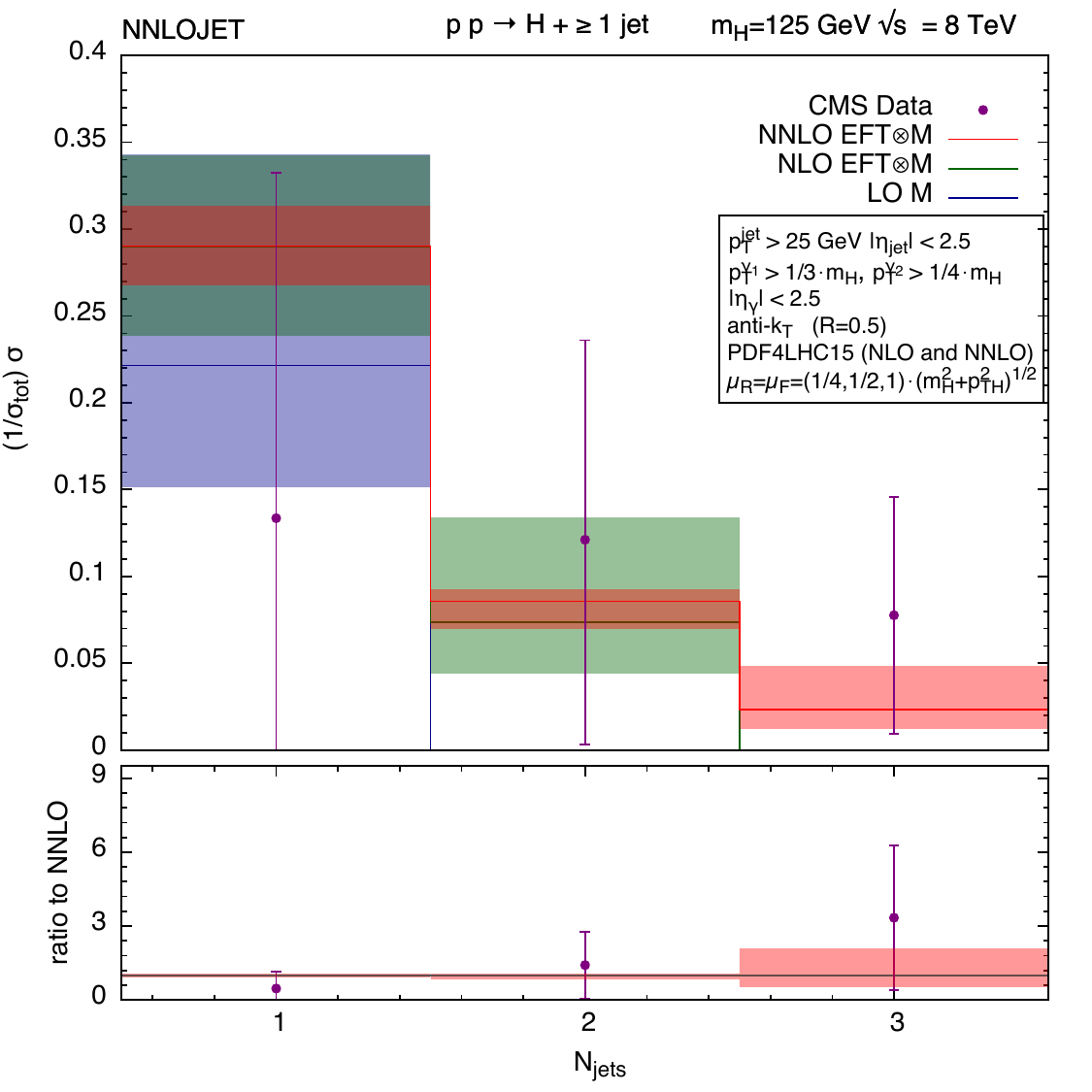}
\caption{Jet multiplicity in Higgs-plus-jet production, normalized to the total fiducial cross section compared to ATLAS~\protect{\cite{atlashpt}} and CMS~\protect{\cite{cmshpt}} data.\label{fig:njetnorm}}
\end{figure}
Figure~\ref{fig:njet} compares the ATLAS and CMS measurements for the jet 
multiplicities with the theoretical \EFTtimes\ predictions up to ${\cal O}(\alpha_s^5)$, which is NNLO for 
Higgs-plus-one-jet final states (but only LO for Higgs-plus-three-jets). For the ATLAS measurement, we observe that the data lies systematically above the theory prediction, while consistency within 
errors is found for the CMS measurement. The same tension between ATLAS data and theory can also be observed in the total fiducial cross section $\sigma_H$ in 
Table~\ref{tab:xsec}. 
By normalizing the jet multiplicities to the corresponding inclusive cross sections, $\sigma_H$, see Figure~\ref{fig:njetnorm}, we see that both ATLAS and CMS data are consistent with the 
theoretical NNLO \EFTtimes\ predictions, which however now display a larger theoretical scale uncertainty (because
the scales are varied independently in numerator and denominator). For exclusive $H+1j$ production, the uncertainty increases from 2\% (for both ATLAS and CMS) for the absolute 
prediction to 11\% (ATLAS) and 8\% (CMS) for the normalized prediction. 
This substantial increase in theory uncertainty suggests that future precision studies of 
Higgs-plus-jet production should preferably be performed on absolute cross sections and distributions.  

\begin{figure}
  \centering
\includegraphics[width=7cm,natwidth=610,natheight=642]{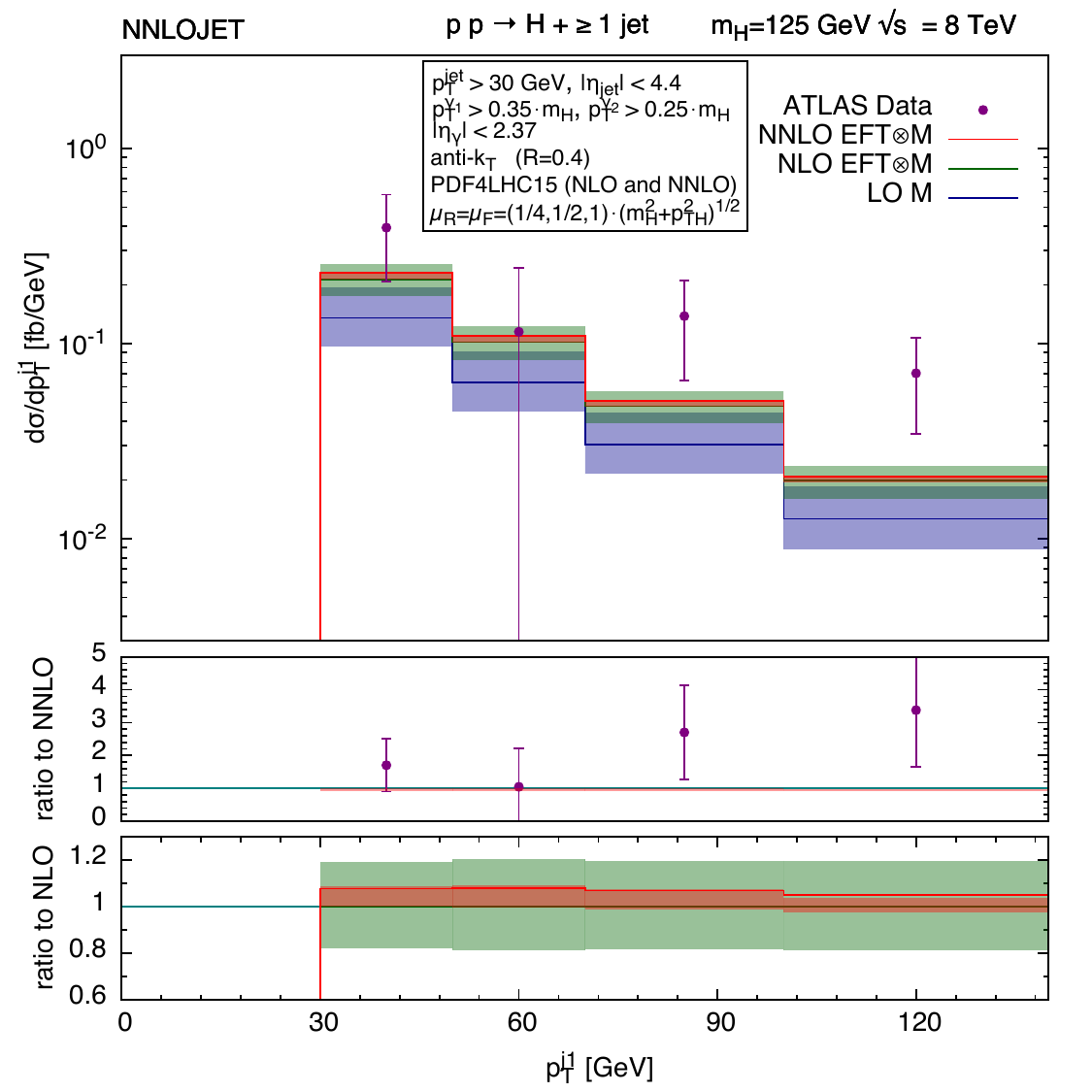}
\includegraphics[width=7cm,natwidth=610,natheight=642]{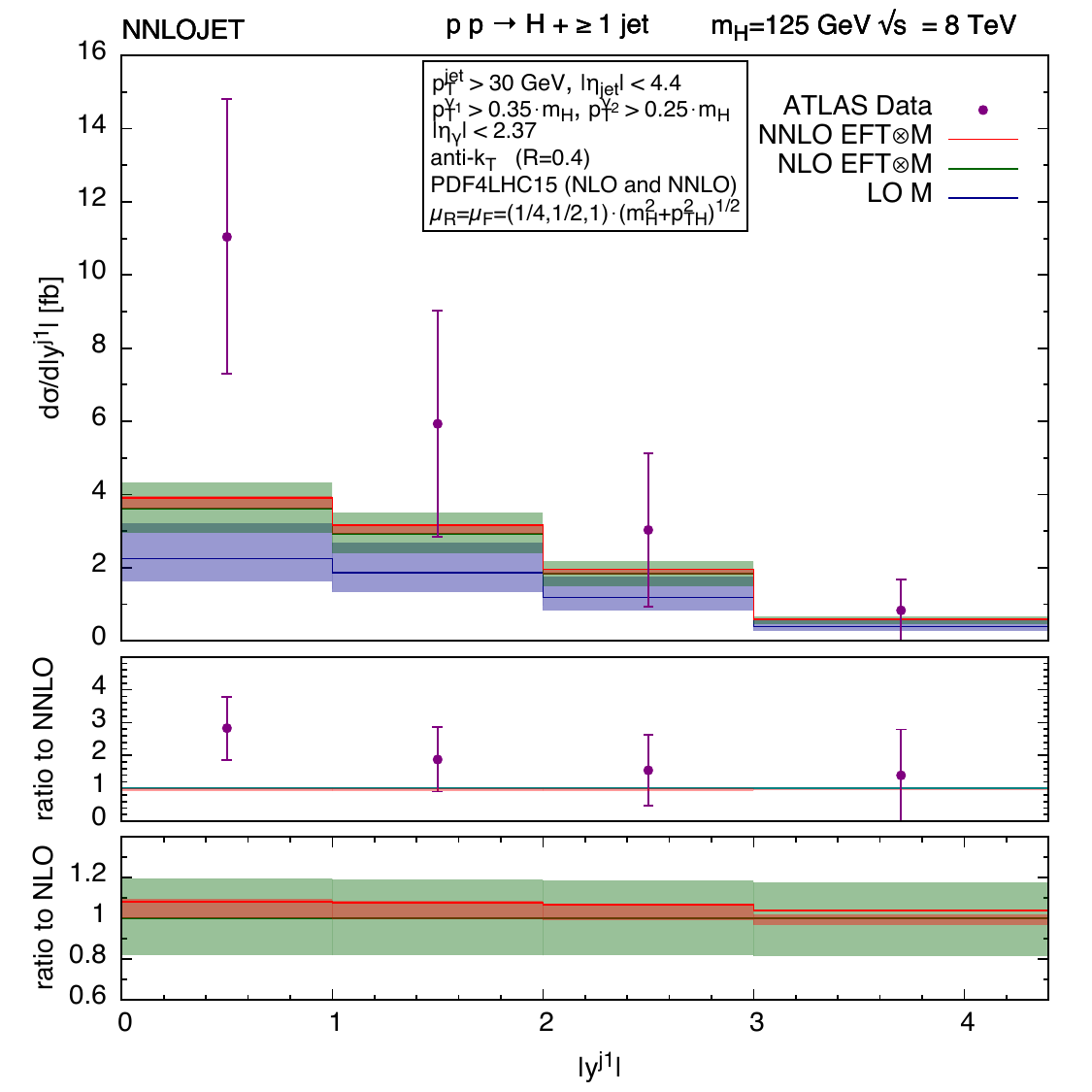}\\
\includegraphics[width=7cm,natwidth=610,natheight=642]{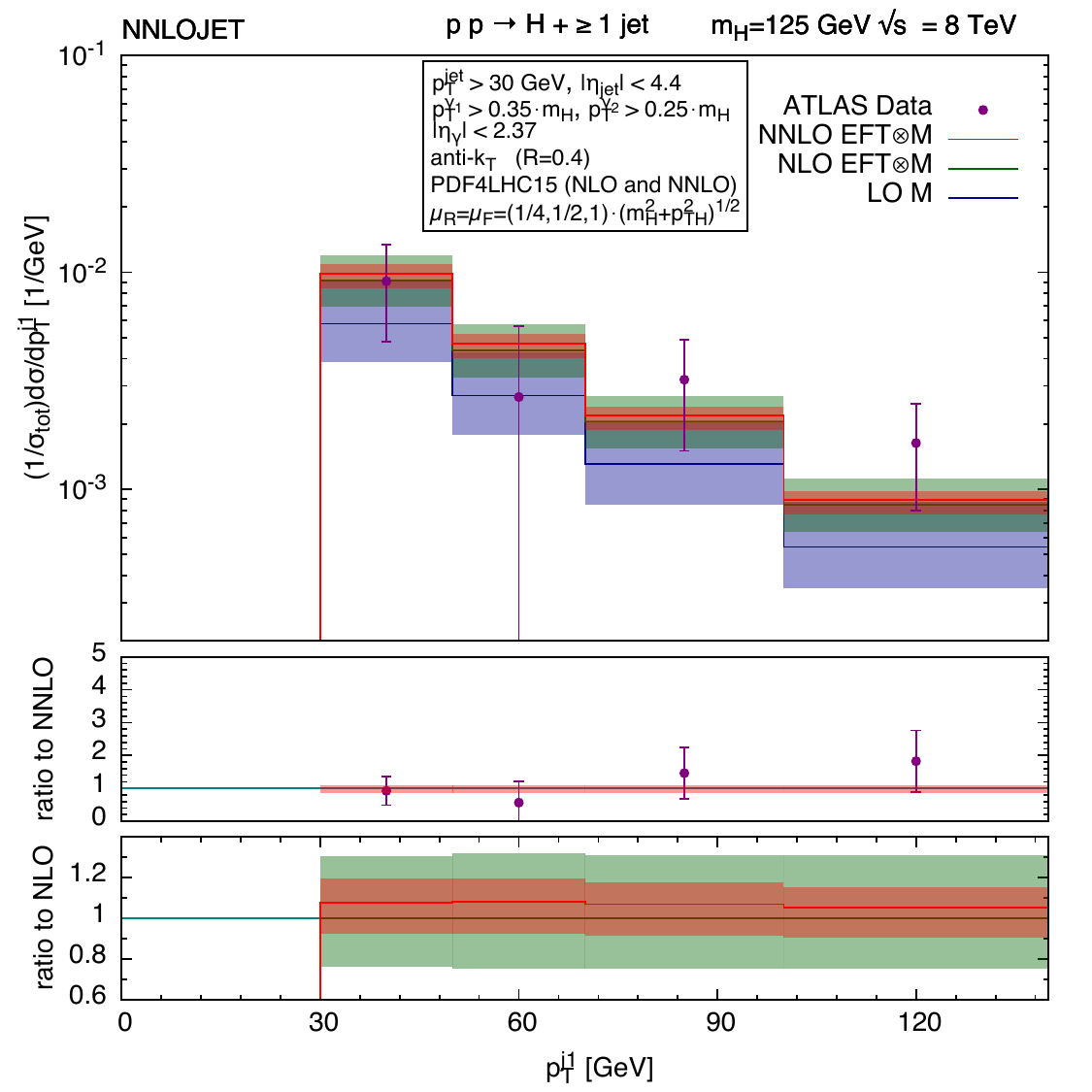}
\includegraphics[width=7cm,natwidth=610,natheight=642]{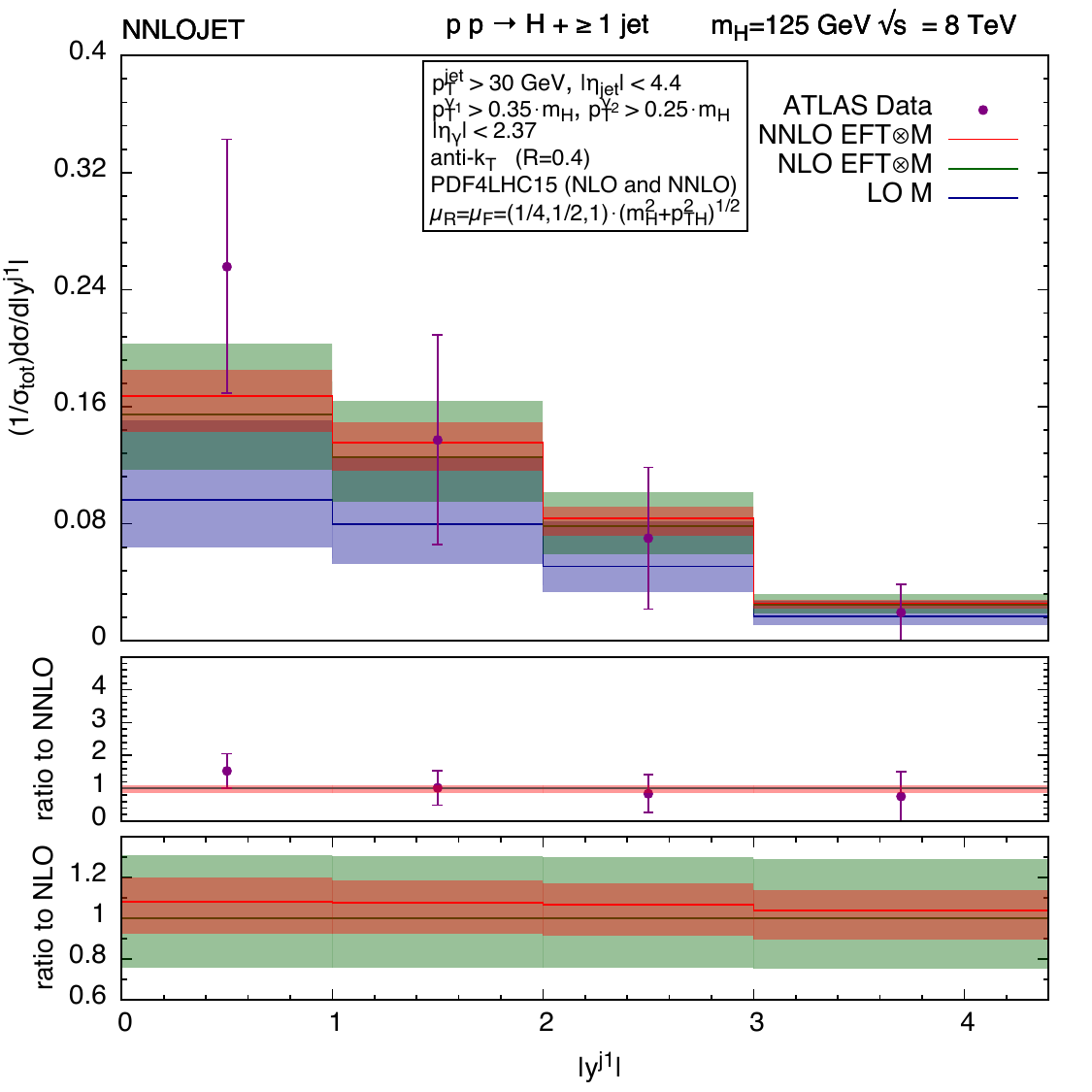}
\caption{Transverse momentum and rapidity distributions of the leading  jet produced in association with a Higgs boson 
 compared to ATLAS data~\protect{\cite{atlashpt}}. Upper panels are absolute cross sections, lower panels normalized 
 to $\sigma_{H}$.\label{fig:atlasjet1}}
\end{figure}
ATLAS and CMS have measured kinematical distributions in Higgs-plus-jet events. When comparing these measurements to the theoretical NNLO  \EFTtimes\ expectations, we consider both absolute and normalized distributions in parallel, in order to discriminate between the description of shapes 
and absolute normalizations. Figure~\ref{fig:atlasjet1} shows the transverse momentum and rapidity distributions for the leading jet, compared to 
ATLAS data~\cite{atlashpt}. 
We observe the NNLO \EFTtimes\ corrections to be significant, typically of the order of +9\% compared to NLO, and to be concentrated in low $p_T^{j1}$ and at central rapidity. The residual uncertainty on the theory prediction is at the level of about 5\%.   
As already observed for the $H+1j$ fraction, the theory prediction falls significantly below the data in absolute normalization. The shape of the data is 
well described by the NNLO theory, as can be seen from the distributions normalized to the fiducial cross section $\sigma_H$ (lower panels in 
Figure~\ref{fig:atlasjet1}). 

\begin{figure}
  \centering
\includegraphics[width=7cm,natwidth=610,natheight=642]{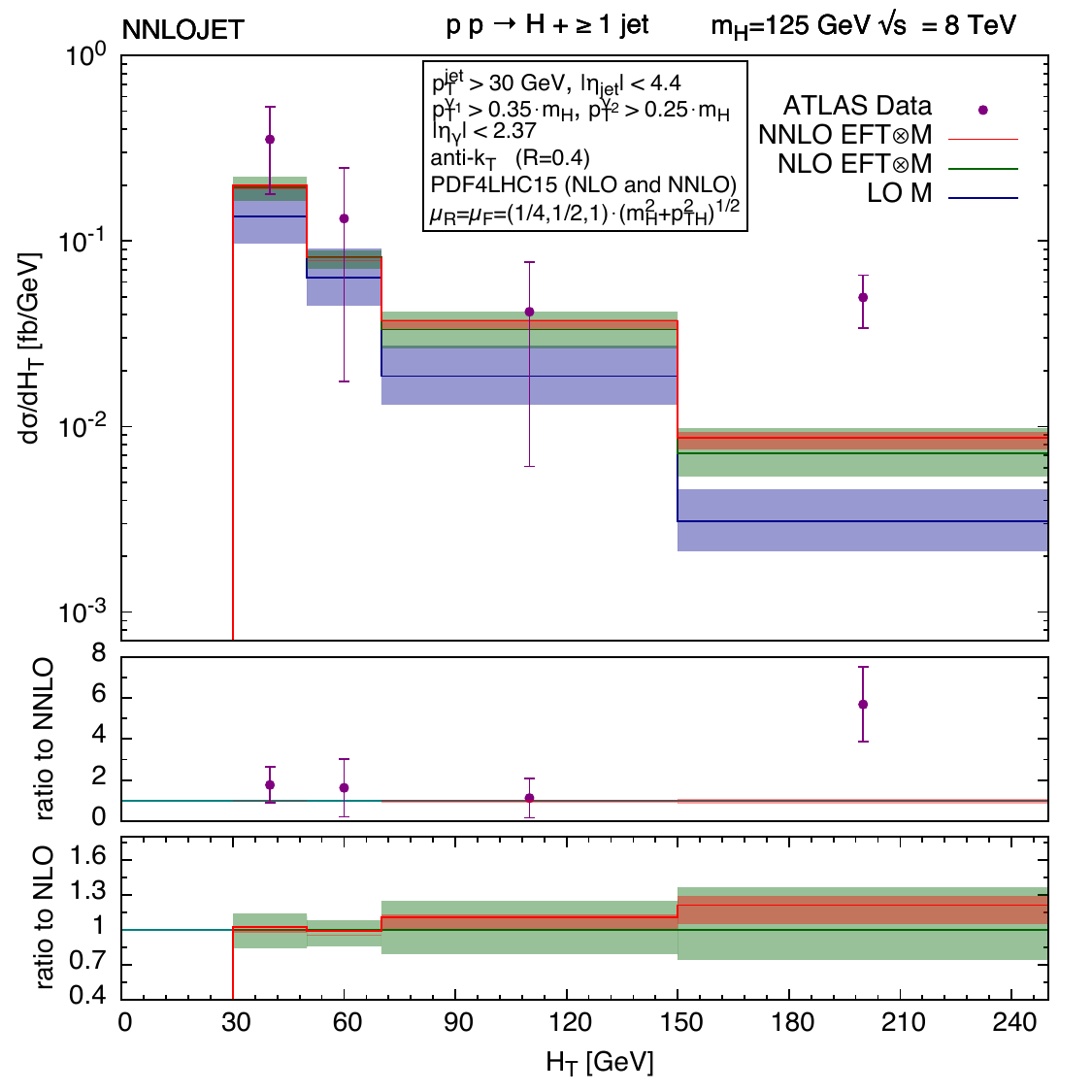}
\includegraphics[width=7cm,natwidth=610,natheight=642]{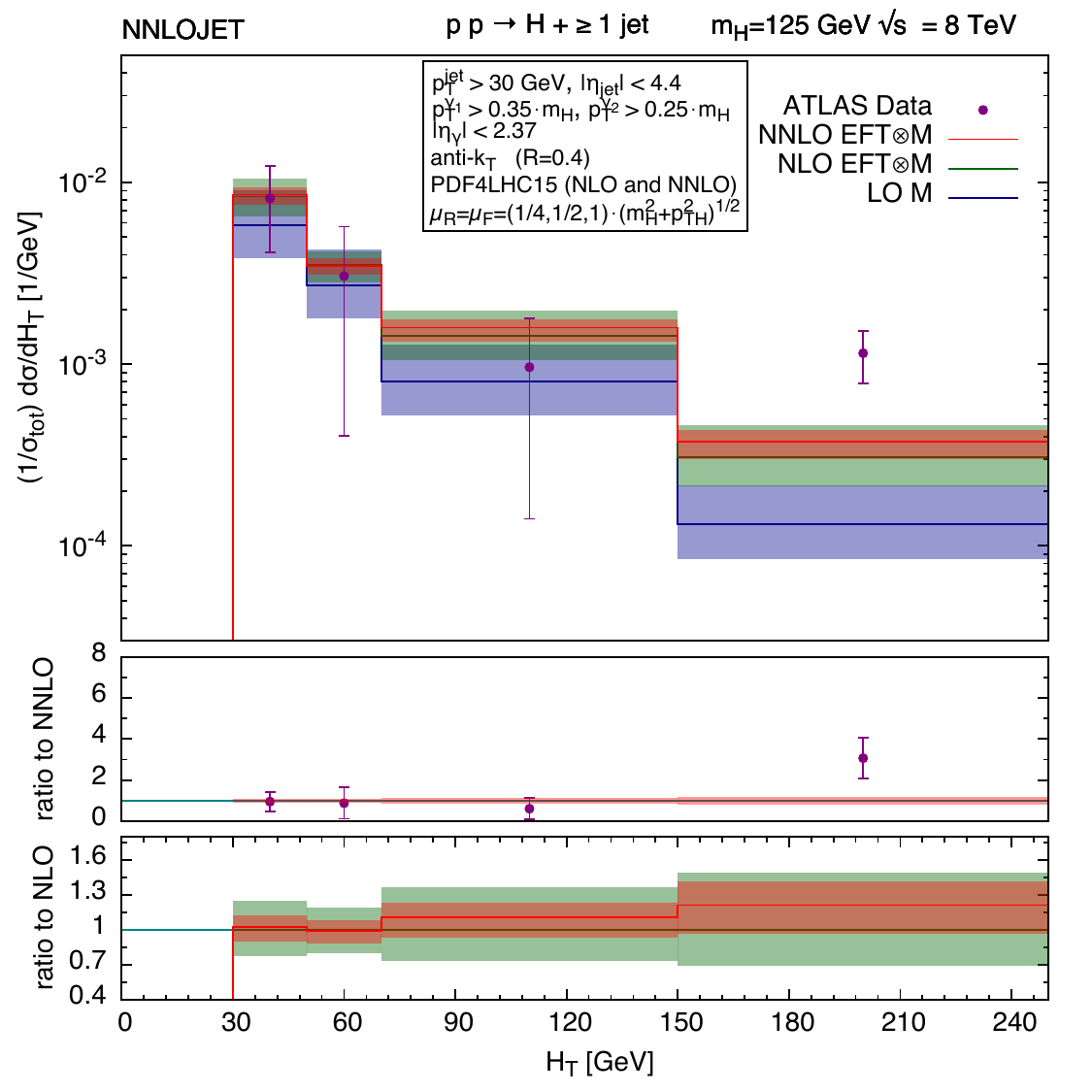}
\caption{Transverse momentum sum of all jets produced  in association with a Higgs boson 
 compared to ATLAS data~\protect{\cite{atlashpt}}. Left panel is the absolute cross sections, right panel normalized 
 to $\sigma_{H}$.\label{fig:atlasjetall}}
\end{figure}
A similar behaviour is also observed for the transverse momentum sum of all jets $H_T$, shown in Figure~\ref{fig:atlasjetall}. The shape of the distribution is 
well-described by NNLO QCD, while the normalization is discrepant by about the same amount as in the fiducial cross section $\sigma_H$.  The NNLO corrections are more significant in the high $H_T$ region at the order of +20\% (compared to NLO).

\begin{figure}
  \centering
\includegraphics[width=7cm,natwidth=610,natheight=642]{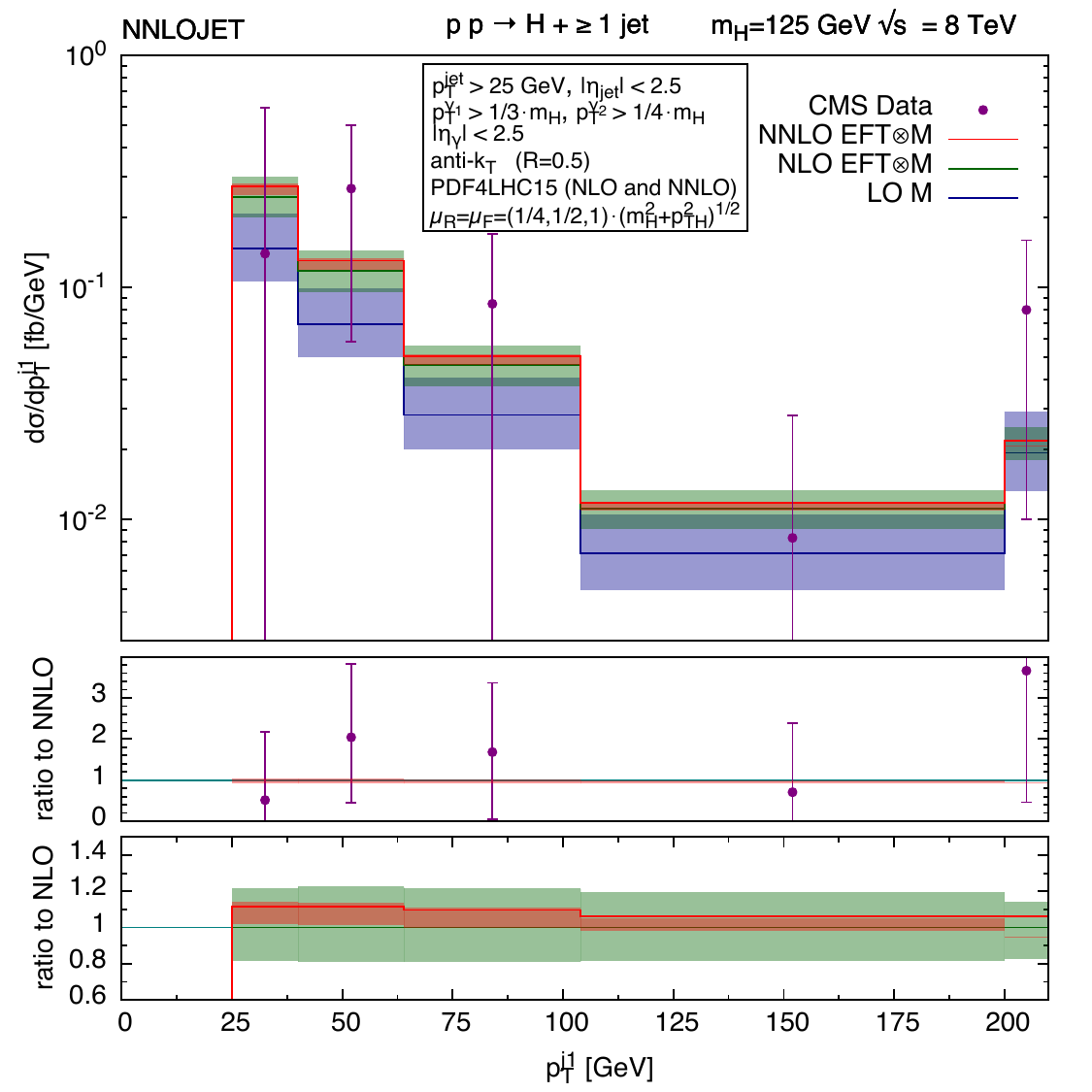}
\includegraphics[width=7cm,natwidth=610,natheight=642]{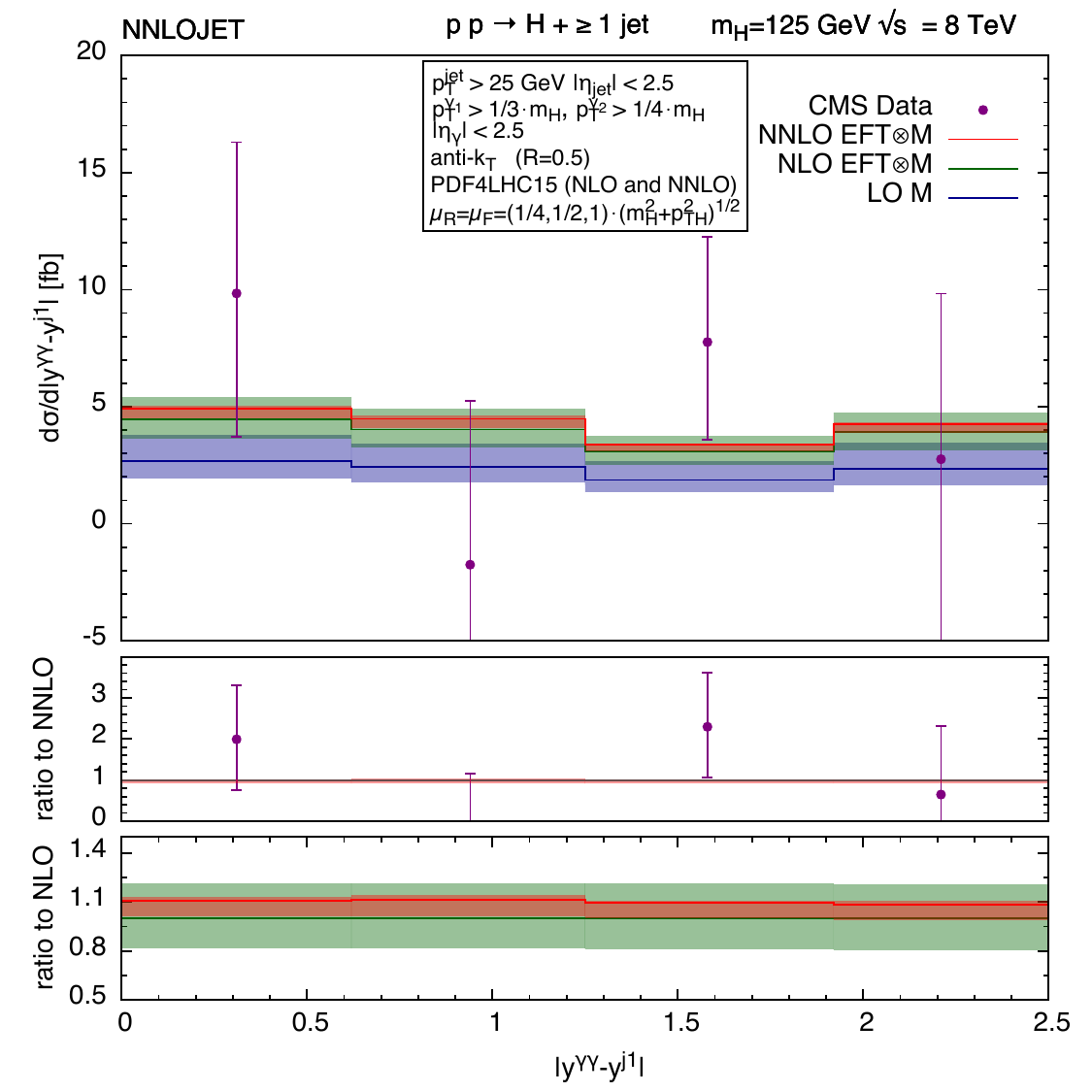}\\
\includegraphics[width=7cm,natwidth=610,natheight=642]{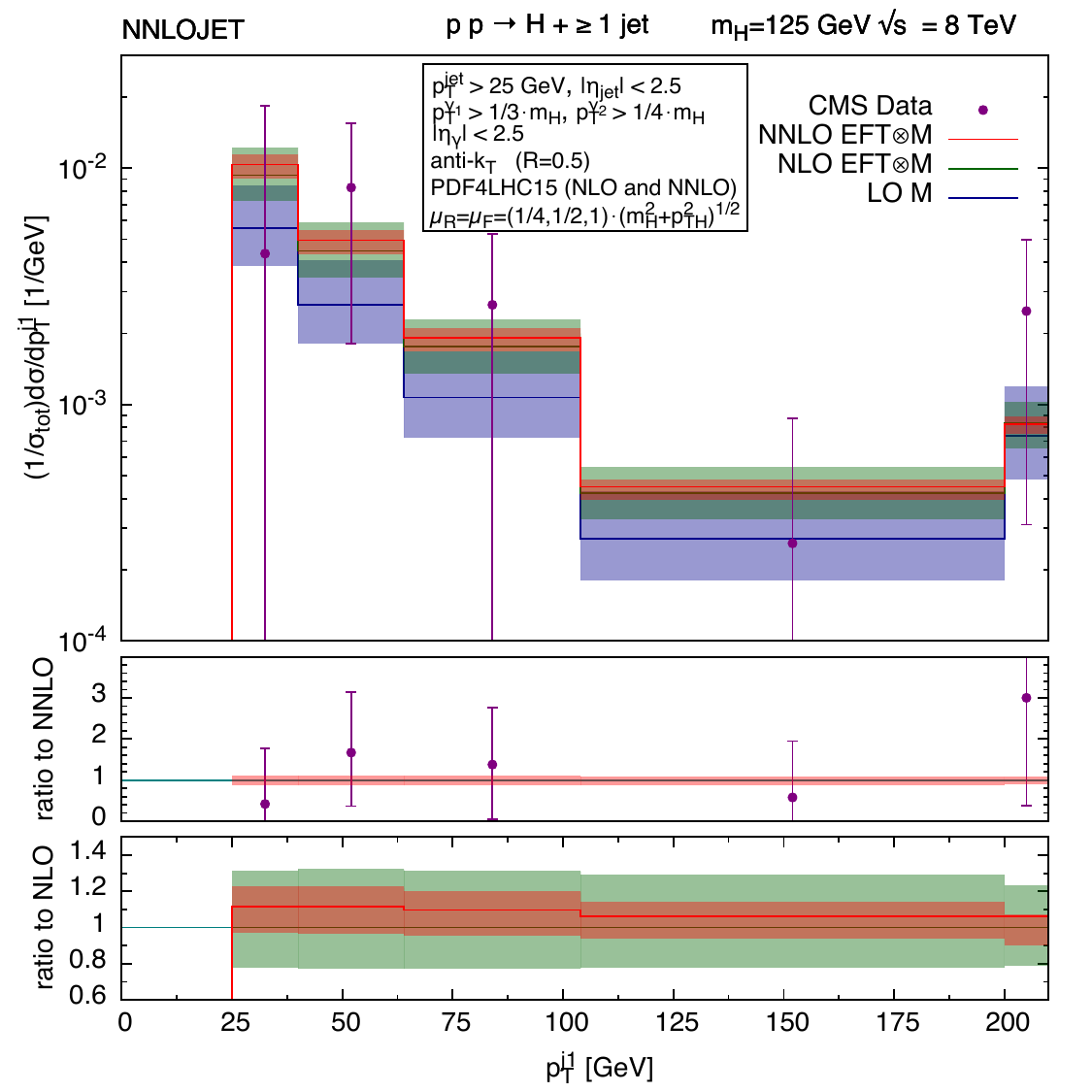}
\includegraphics[width=7cm,natwidth=610,natheight=642]{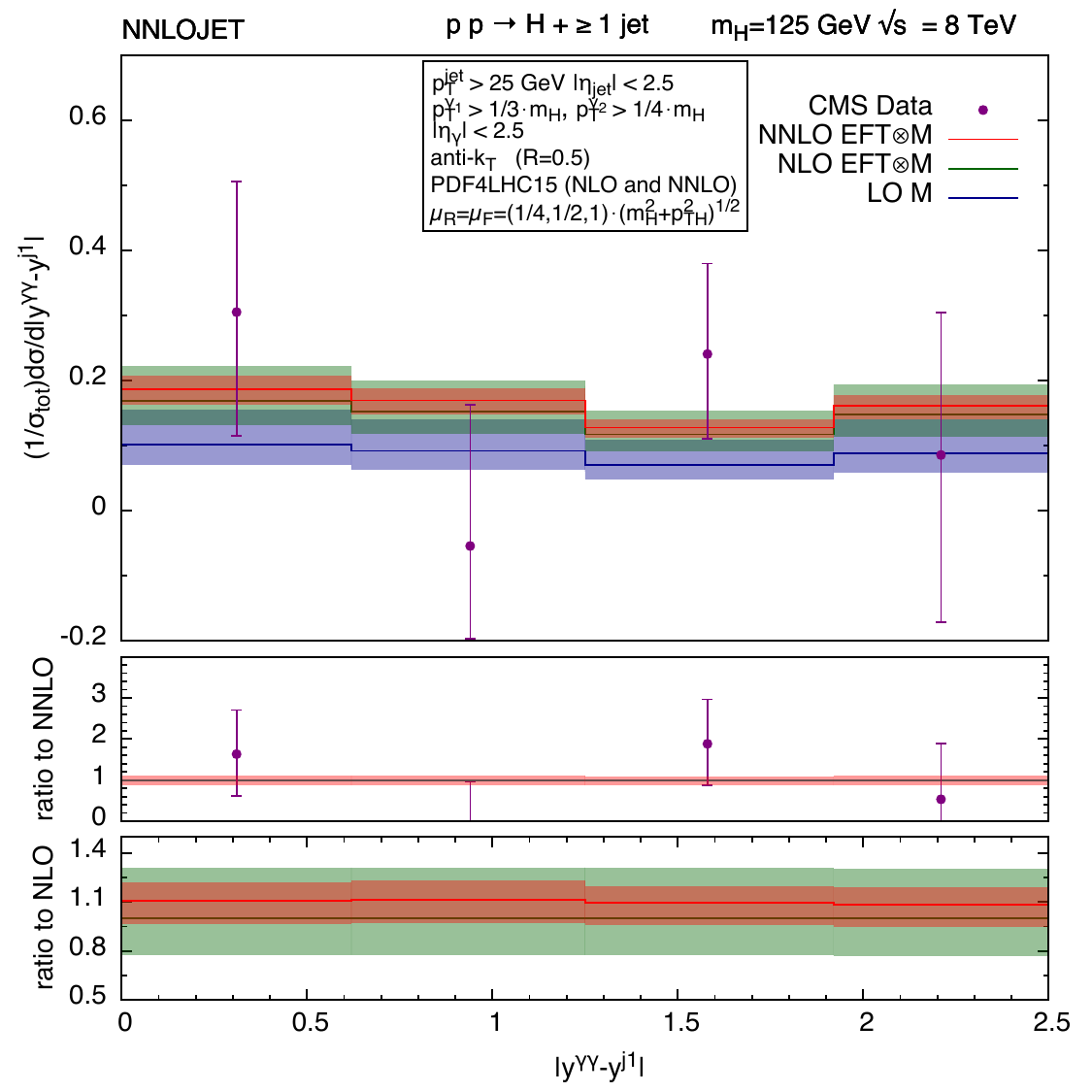}
\caption{Transverse momentum distribution and 
rapidity correlation of the leading jet produced in association with a Higgs boson 
 compared to CMS data~\protect{\cite{cmshpt}} as absolute cross sections (upper row) and normalized to $\sigma_{H+j}$.\label{fig:cmsjet1}}
\end{figure}
The CMS experiment has measured the transverse momentum distribution of the leading jet in Higgs-plus-jet events and the rapidity separation between 
the Higgs boson and the leading jet. We compare these measurements to our NNLO  \EFTtimes\  predictions in Figure~\ref{fig:cmsjet1}. The last bin in Figure~\ref{fig:cmsjet1} contains the overflow beyond the right edge for both experiment data and theory predictions. We see that the 
NNLO corrections are largest at low transverse momentum but are generally uniform in rapidity separation. The NNLO corrections are  
somewhat larger than for the ATLAS cuts at the order of +11\% compared to NLO and find the NNLO scale uncertainty to be 
about 6\%. The larger NNLO corrections may be related to the fact that CMS uses a larger jet radius than ATLAS. The absolute normalization of 
the CMS data is already well described by NNLO QCD, such that normalization to the total fiducial cross sections does not modify the quantitative 
comparison between data and theory. 

In this section, we have presented NNLO QCD results for fiducial cross sections in Higgs-plus-jet production in the diphoton decay mode taking the LO mass effects into account according to the \EFTtimes\ precription. Our results 
were obtained with the \NNLOJET code, which is based on the antenna subtraction method. Overall, we 
observe the corrections to be positive and moderate in size. The NNLO predictions are typically at the upper edge of the NLO scale variation interval, 
and come themselves with a residual theory uncertainty of around 5\%. We observe that the ATLAS measurements~\cite{atlashpt} 
are well-described in shape, but not in normalization, a feature that also persists to the same 
magnitude in the total fiducial cross section, which is inclusive in the number of
jets. 
Besides the absolute distributions, we therefore also considered distributions normalized 
to the total fiducial cross section. In these, uncertainties related to the overall luminosity and the reconstruction efficiency largely cancel out, such that 
normalized distributions are often measured more reliably. 
We observe the theory uncertainty on the distributions to increase after normalization, which is a 
direct consequence of considering independent scale variations on numerator and denominator. For this reason, they appear to be less well suited for 
precision phenomenology than the absolute measurements. 

\subsection{Higgs boson transverse momentum distribution}
\label{sec:hpt}

\begin{figure}
  \centering
\includegraphics[width=7cm,natwidth=610,natheight=642]{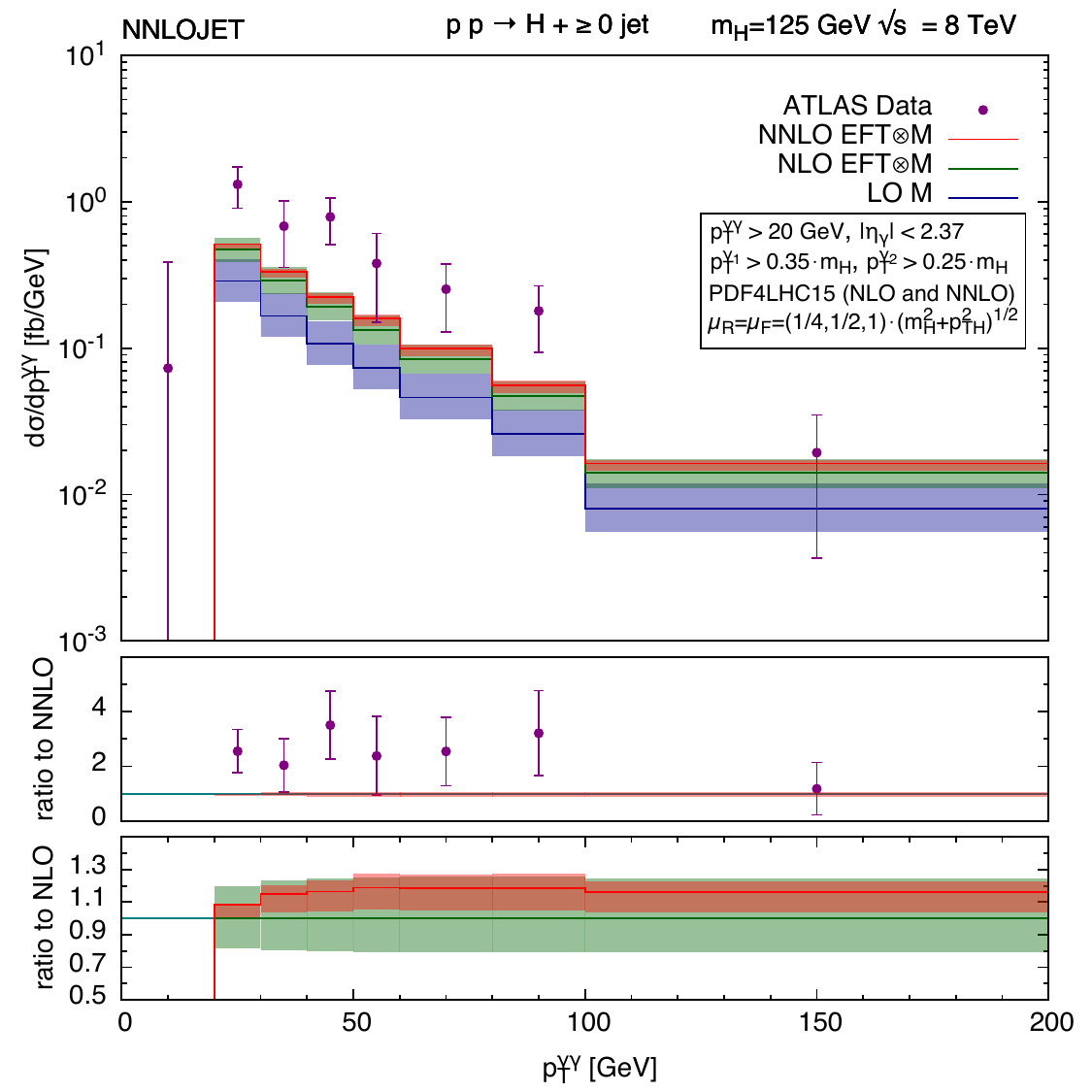}
\includegraphics[width=7cm,natwidth=610,natheight=642]{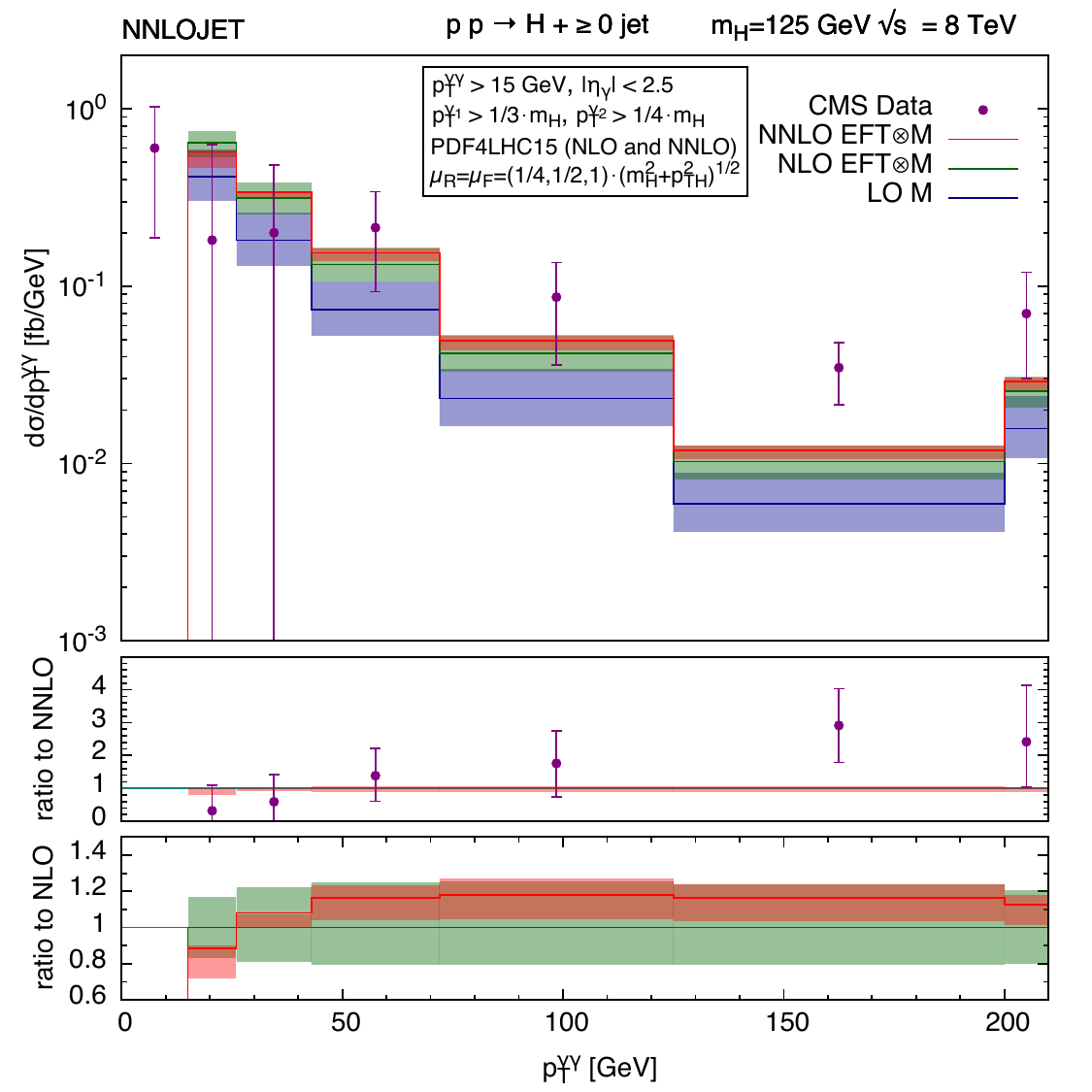}\\
\includegraphics[width=7cm,natwidth=610,natheight=642]{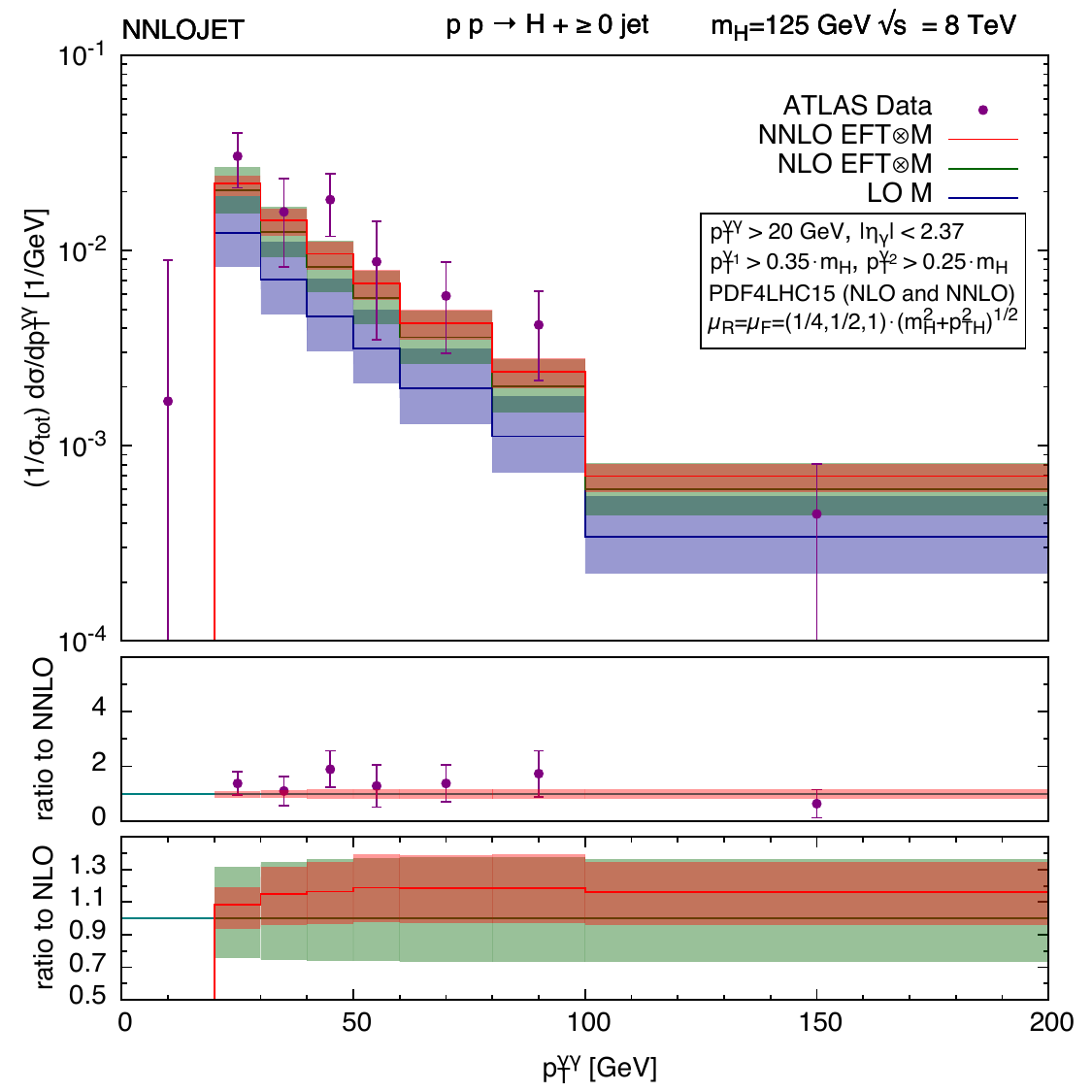}
\includegraphics[width=7cm,natwidth=610,natheight=642]{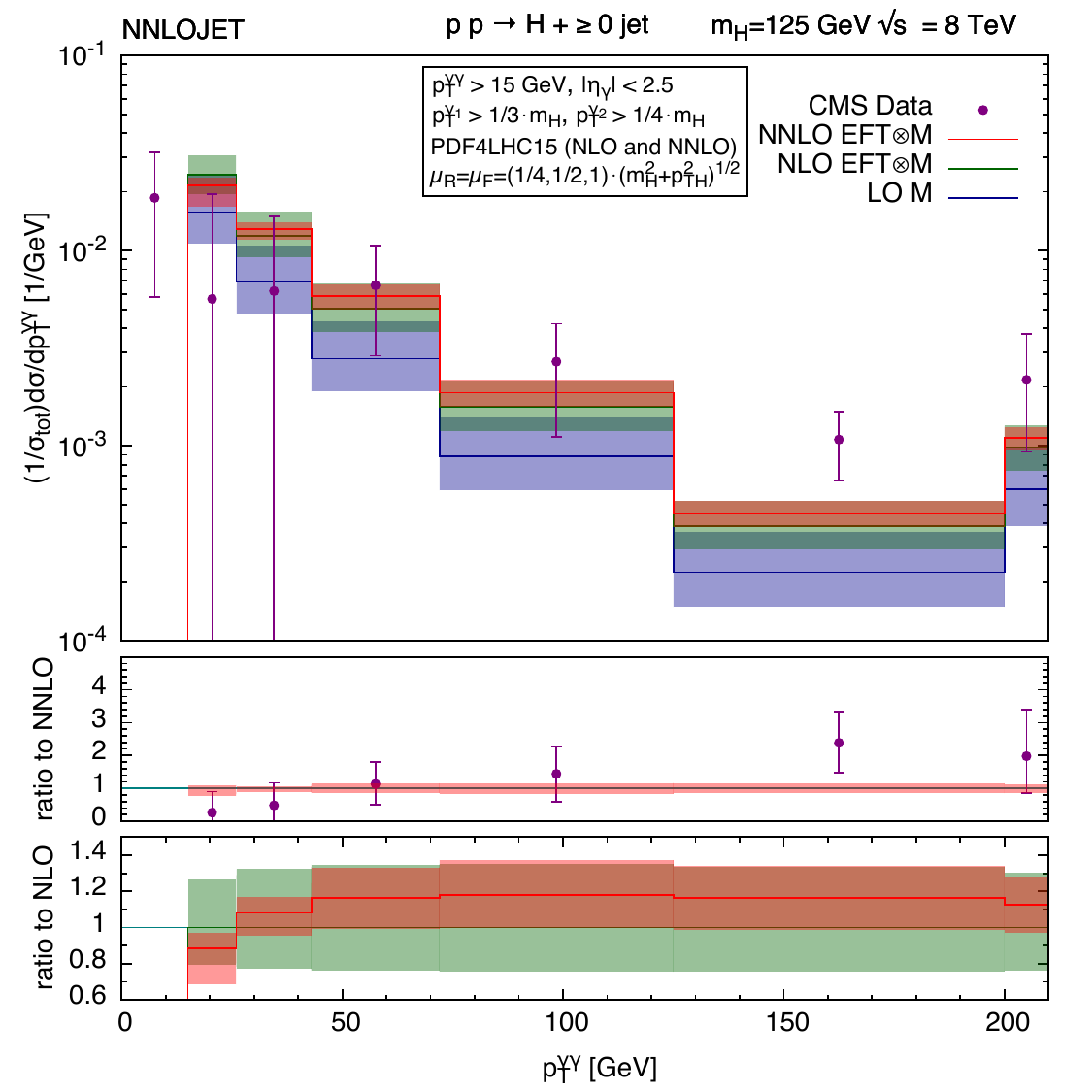}
\caption{Transverse momentum distribution of the Higgs boson compared to data from ATLAS~\protect{\cite{atlashpt}} and CMS~\protect{\cite{cmshpt}}.
Upper panels are absolute cross sections, lower panels normalized 
 to $\sigma_{H}$.
\label{fig:pth}}
\end{figure}
Closely related to Higgs-plus-jet final states is the production of a Higgs boson at finite transverse momentum. Since the transverse momentum 
of the Higgs boson is 
generated by the recoil against a parton, this process receives exactly the same parton-level contributions as Higgs-plus-jet production. 
Theoretical predictions for it can thus be derived from a calculation of Higgs-plus-jet final states by replacing the jet reconstruction criterion by 
a lower cut on the transverse momentum of the Higgs boson. In the limit of vanishing transverse momentum, this calculation is infrared 
divergent. The lower transverse momentum cut can therefore not be taken too small in order to avoid instabilities in the evaluation of different ingredients. 
In our evaluation, we use the \NNLOJET code with $p_T^H > 15$~GeV for the CMS cuts and  $p_T^H > 20$~GeV for the ATLAS cuts, each time 
corresponding to lower edge of the second bin in the measured distributions~\cite{atlashpt,cmshpt}. The first bin in these distributions contains the infrared divergent contribution from the fixed-order process, and can be described reliably only if virtual contributions from Higgs boson production at vanishing
 $p_T^H$ are included as well. This can be accomplished by the $H+0j$ fixed-order process with an 
upper veto on the Higgs boson transverse momentum, or by using a resummed formulation. As a consequence, the fixed order ${\cal O}(\alpha_s^5)$ is 
NNLO for finite $p_T^H$, but corresponds to N3LO for the first bin. At N3LO, only the computation of 
the total cross section for Higgs boson production~\cite{ggHn3lo} and the  exclusive $H+0j$ cross section (jet veto cross section, \cite{hjveto}) 
have been completed so far. Results for fiducial cross sections are not yet available. 

Figure~\ref{fig:pth} compares the Higgs boson transverse momentum distributions measured by ATLAS~\cite{atlashpt} and CMS~\cite{cmshpt} 
to our NNLO  \EFTtimes\  predictions. The last bin of the CMS measurement and theory prediction (left panels) contains the overflow. The NNLO corrections lead to a uniform increase of the theoretical prediction by about 15\% compared to NLO, overlapping with the upper edge of the NLO uncertainty band. The remaining theory scale uncertainty is at the level of 8\% for the 
unnormalized transverse momentum distribution. 
As for the Higgs-plus-jet production discussed in the previous section, we observe that the shape of the data is 
well-described for both experiments, while the normalization is reproduced only for CMS, while the ATLAS data are systematically above
the theoretical prediction. Normalizing to the total inclusive cross section $\sigma_H$ reconciles data and theory, however at the expense of an 
increase of the theory scale uncertainty to 15\%.

\subsection{Comparison with preliminary 13 TeV data}
\label{sec:ichep}
 
 \begin{figure}[t]
  \centering
\includegraphics[width=7cm,natwidth=610,natheight=642]{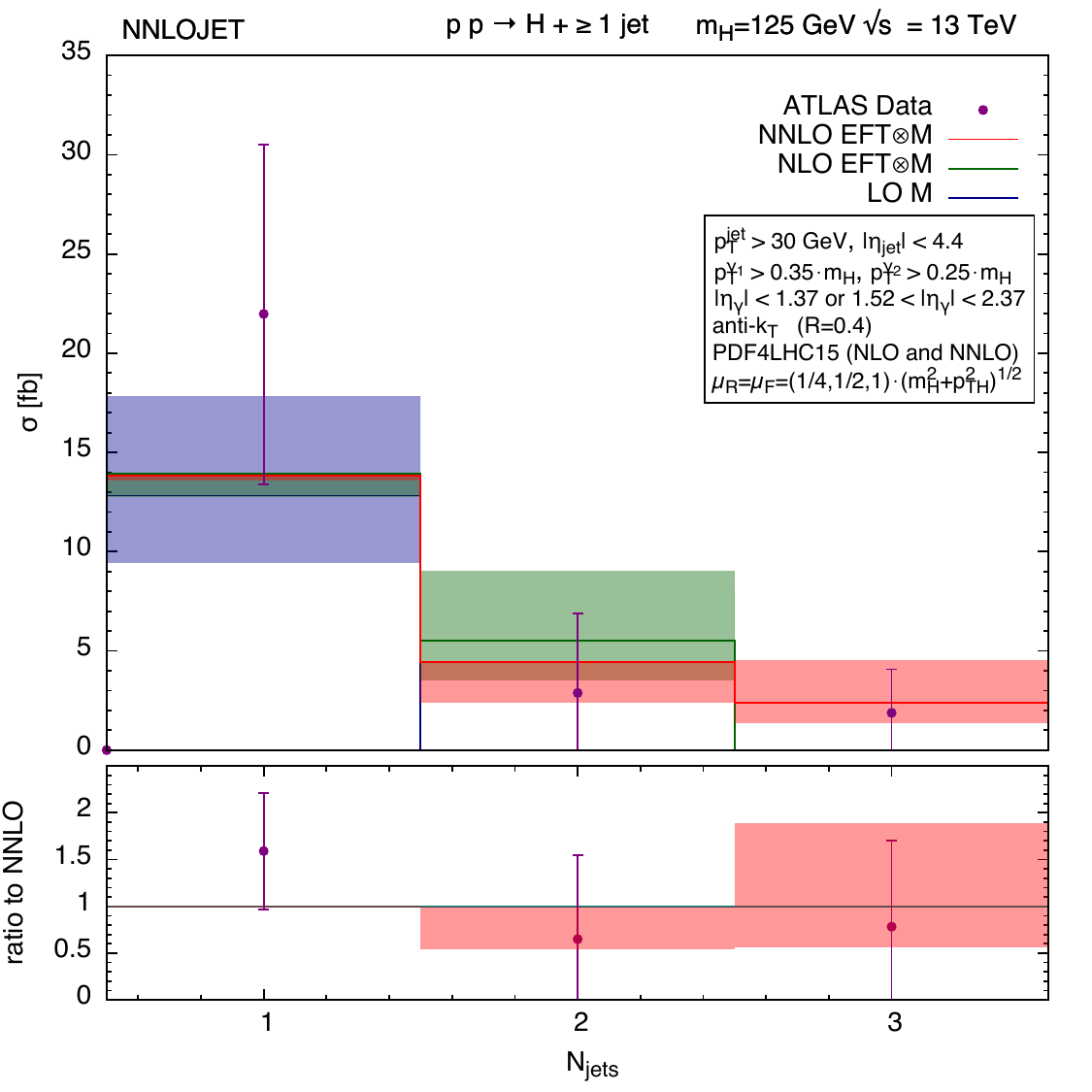}
\includegraphics[width=7cm,natwidth=610,natheight=642]{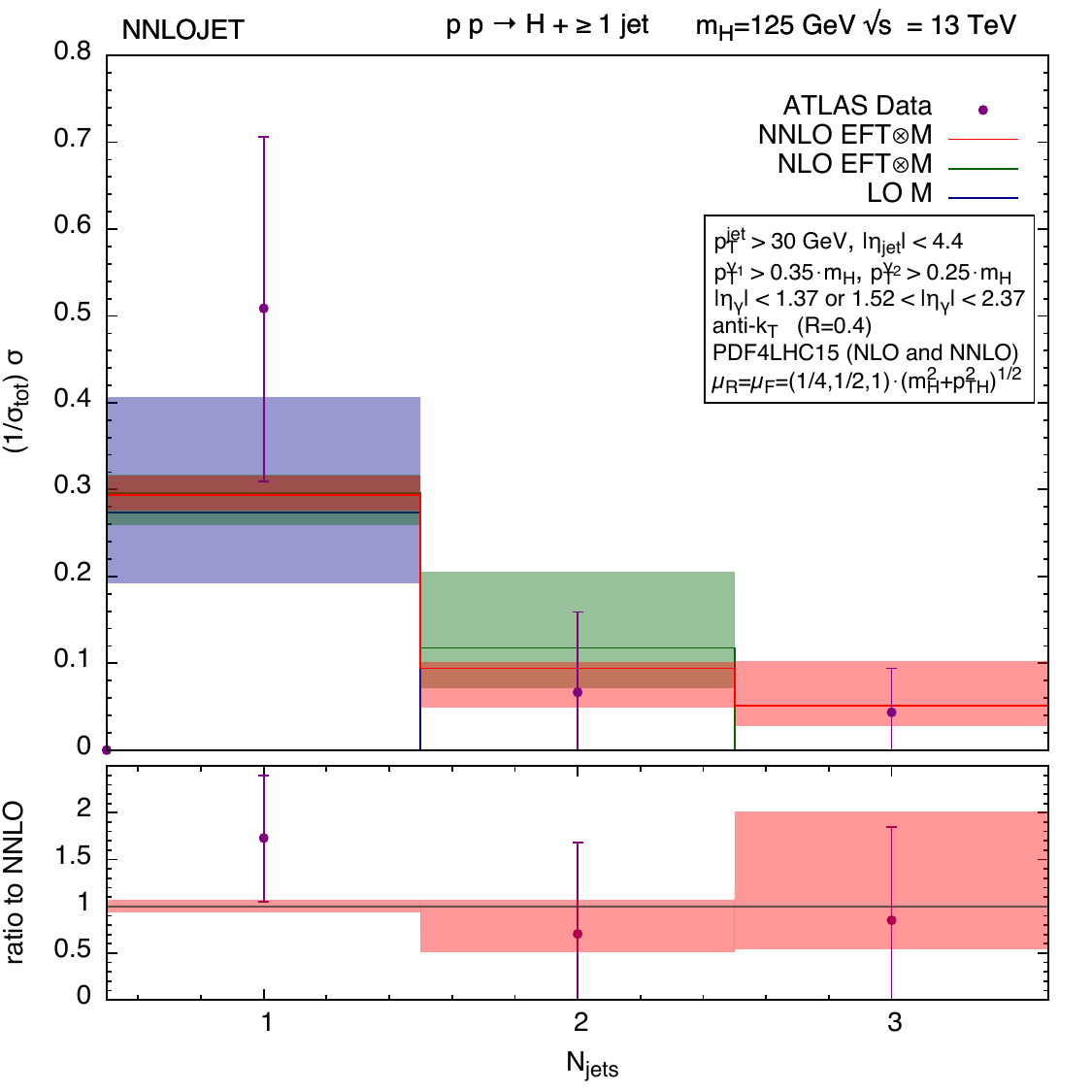}
\caption{Jet multiplicity in Higgs-plus-jet production compared to preliminary 13 TeV ATLAS data~\protect{\cite{atlasICHEP}}. Left panel is the absolute cross section, right panel is normalized to $\sigma_H$.\label{fig:njetICHEP}}
\end{figure}
Recently, the ATLAS Collaboration have presented preliminary measurements of Higgs boson properties in the diphoton channel using 13.3 fb$^{-1}$ of data from LHC Run 2 at 13 TeV~\cite{atlasICHEP}. This dataset is of comparable statistical quality to the 8 TeV measurements discussed above: the lower 
integrated luminosity is compensated for by the increase in the Higgs production cross section at the higher collider energy. 
The experimental event selection is identical to the ATLAS analysis at 8 TeV (see Table~\ref{tab:cuts}), with an additional criterion of 
excluding photons 
in the pseudorapidity range $1.37 < |\eta|<1.52$. The measured fiducial cross section~\cite{atlasICHEP} of 
$\sigma_{H,{\rm exp}} = 43.2\pm 14.9 \mbox{(stat)} \pm 4.9 \mbox{(sys)}$~ fb is in good agreement with our NNLO prediction
$\sigma_{H,NNLO}^{EFT \otimes M} = 47.0^{+2.20}_{-3.04}$~ fb. 

The measured jet multiplicity at 13 TeV, Figure~\ref{fig:njetICHEP}, is well described within its still substantial statistical errors 
 by the theoretical prediction. On the absolute magnitude of the $H$+jet cross sections, the agreement is considerably better than for the 
 ATLAS measurement at 8~TeV, Figure~\ref{fig:njet}. Normalising the data and theory predictions to the fiducial cross sections does not 
 alter this agreement, but leads to an increase in the theory uncertainty. 
 \begin{figure}
  \centering
\includegraphics[width=7cm,natwidth=610,natheight=642]{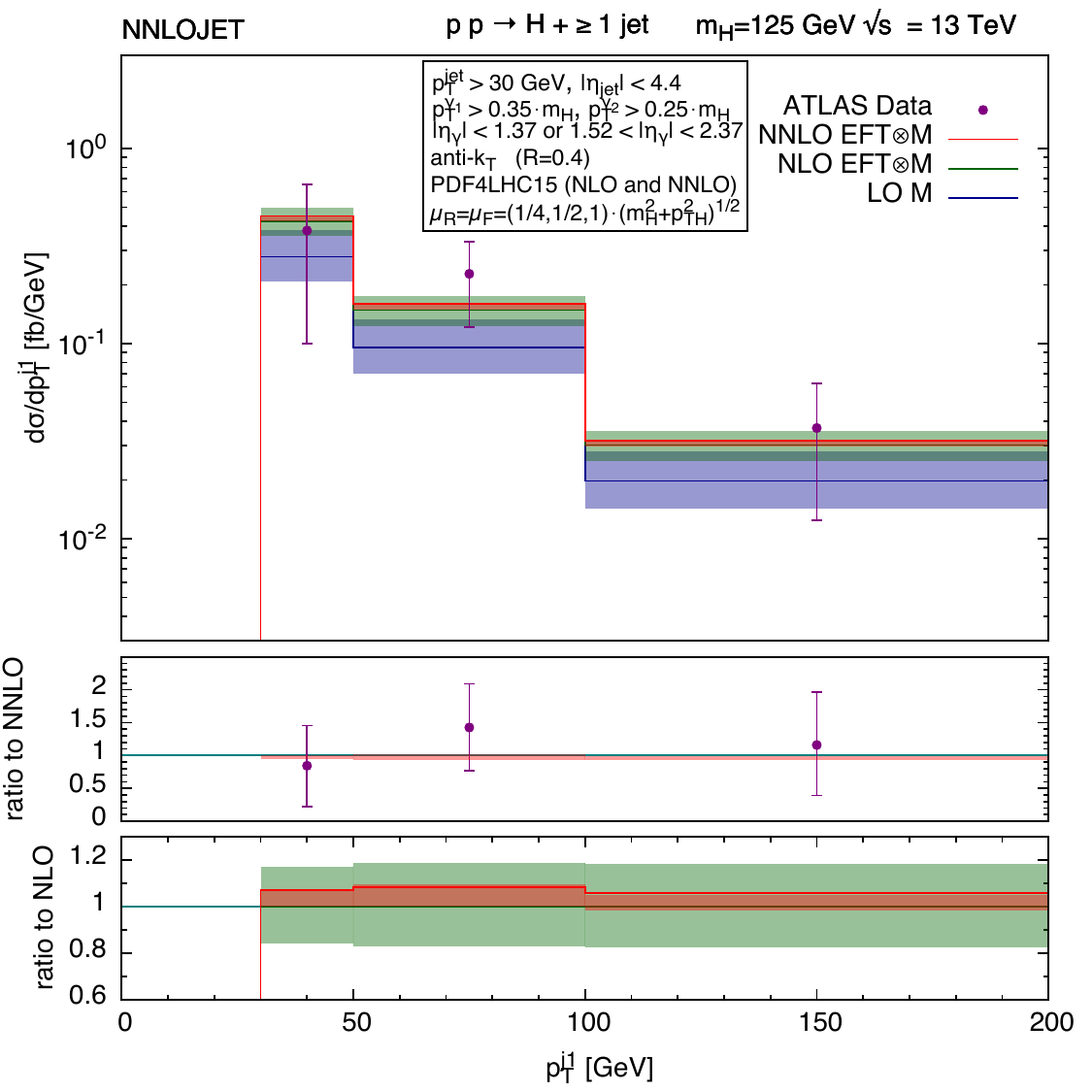}
\includegraphics[width=7cm,natwidth=610,natheight=642]{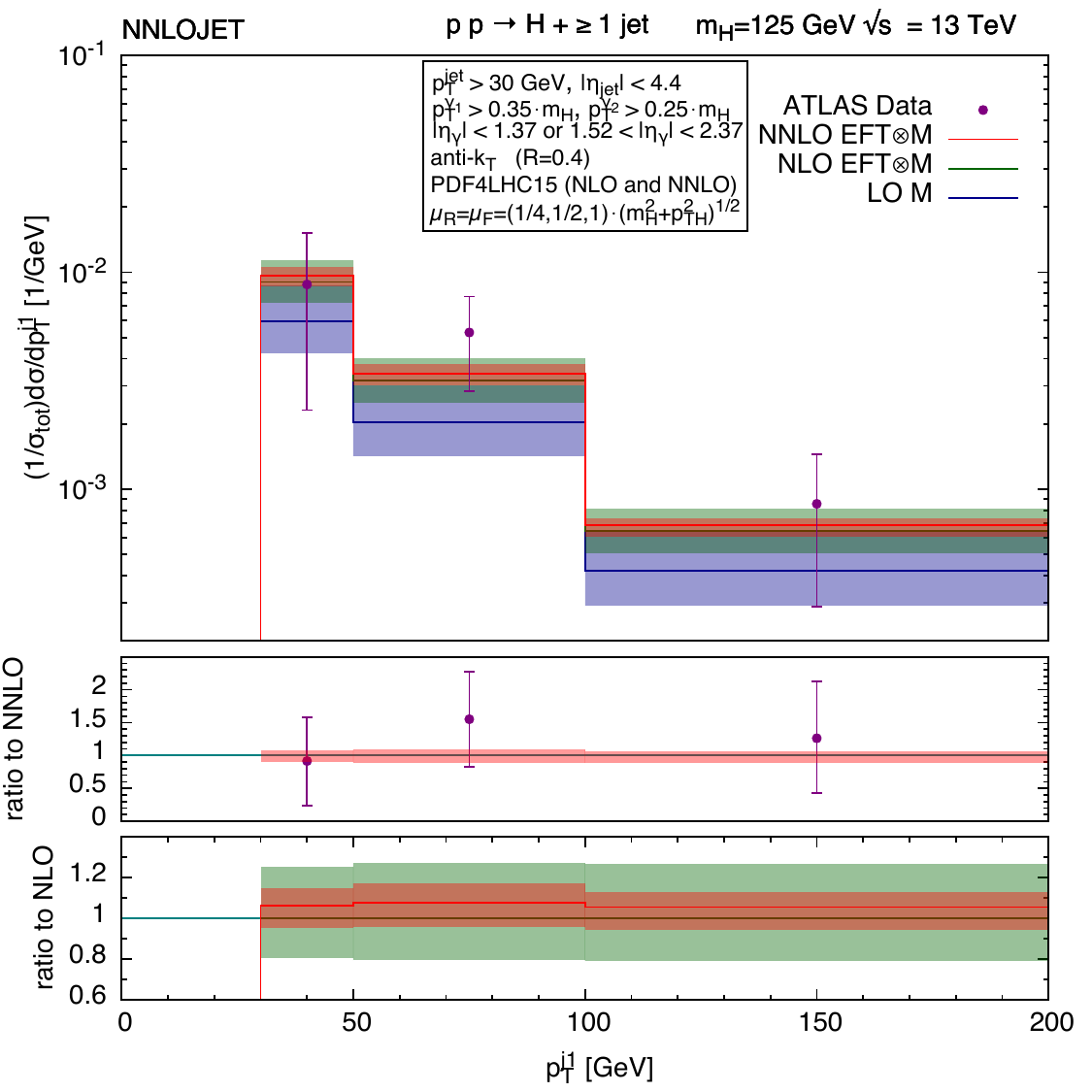}
\caption{Transverse momentum distributions of the leading  jet produced in association with a Higgs boson 
 compared to preliminary 13 TeV ATLAS data~\protect{\cite{atlasICHEP}}. Left panel is the absolute cross section, right panel is normalized to $\sigma_H$.
\label{fig:atlasjet1ICHEP}}
\end{figure}

\begin{figure}
  \centering
\includegraphics[width=7cm,natwidth=610,natheight=642]{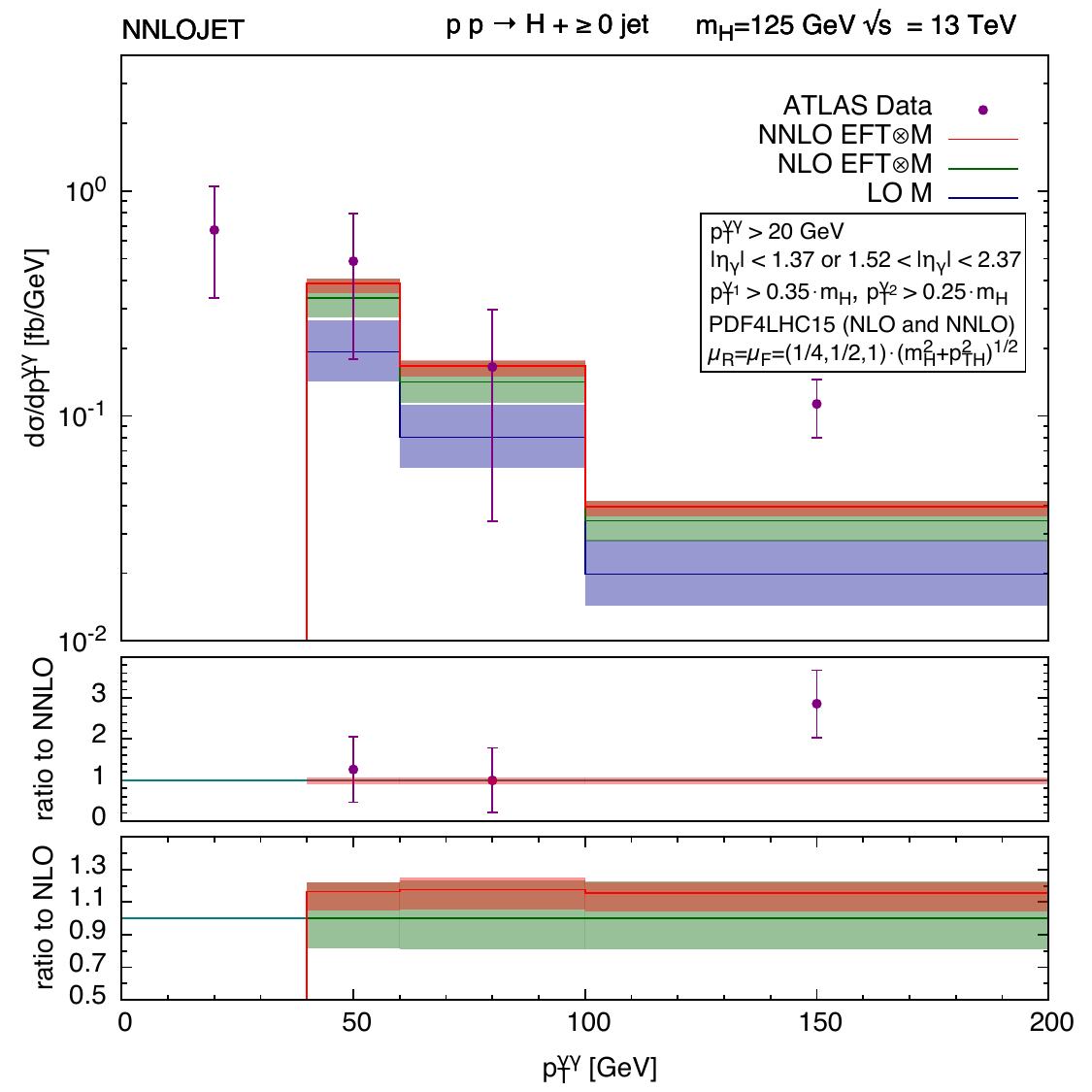}
\includegraphics[width=7cm,natwidth=610,natheight=642]{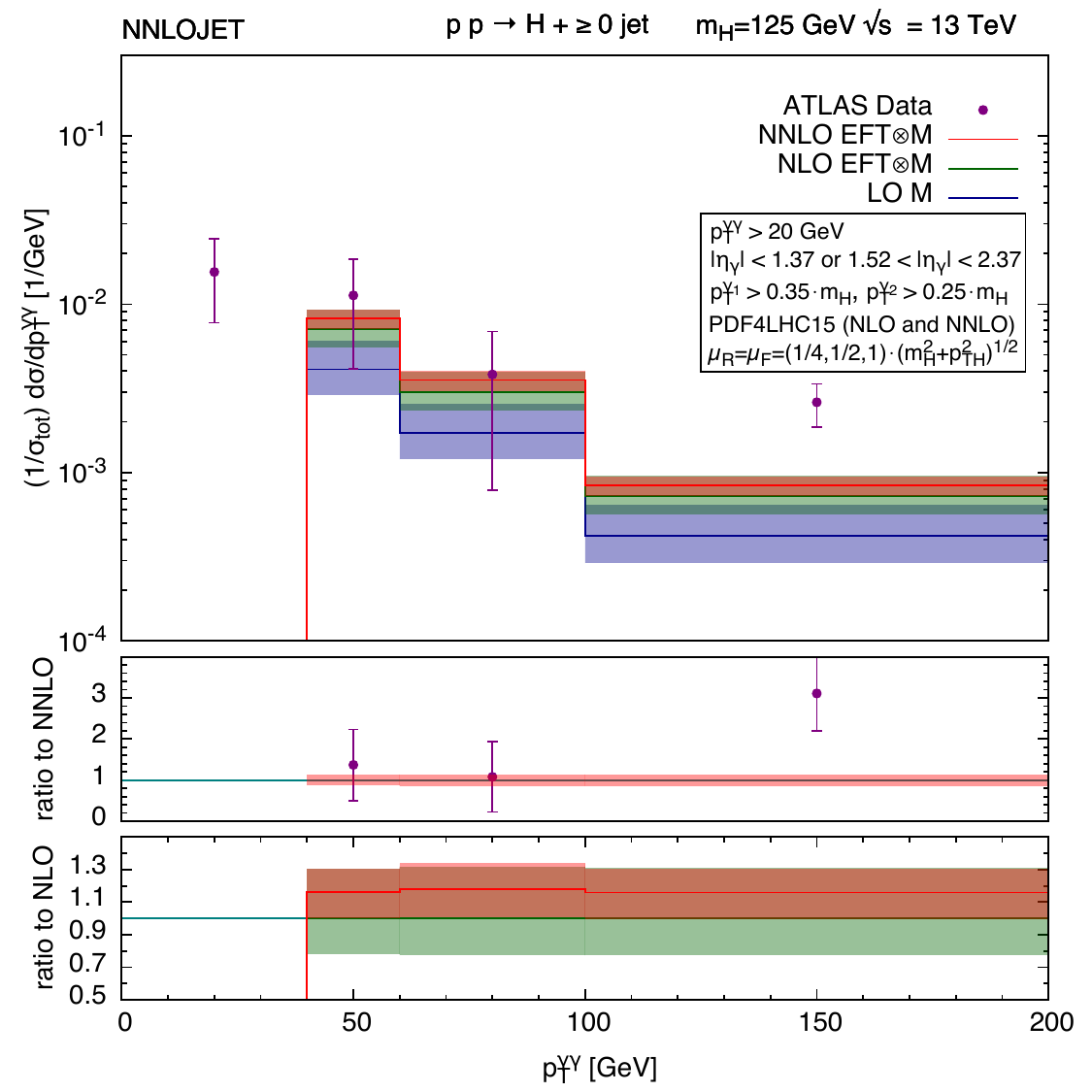}
\caption{Transverse momentum distribution of the Higgs boson compared to preliminary 13 TeV ATLAS~\protect{\cite{atlasICHEP}}. Left panel is the absolute cross section, right panel is normalized to $\sigma_H$.
\label{fig:pthICHEP}}
\end{figure}

The transverse momentum distribution of the leading jet, Figure~\ref{fig:atlasjet1ICHEP}, and of the Higgs boson~\ref{fig:pthICHEP} 
 were both measured by ATLAS up to transverse momenta of 200~GeV. The measurements agree well with our NNLO predictions in 
 shape and normalisation already for the absolute distributions, except for the highest bin in the Higgs transverse momentum distribution, which
 is measured to be about two standard deviations above the theory prediction. As already observed for the jet multiplicity
 at 13 TeV, this quantitative
 agreement persists for the normalised distributions. 

The currently ongoing Run 2 of the LHC will produce a dataset at 13 and 14 TeV corresponding to about 25 times the integrated luminosity 
of the data analysed by ATLAS for the preliminary study~\cite{atlasICHEP} discussed in this section. 

\section {Higgs boson production at large transverse momentum}
\label{sec:hpt2}

Although not yet  very precise, the ATLAS and CMS measurements of the Higgs boson transverse momentum distribution at 
8 TeV~\cite{atlashpt,cmshpt}, as well as the preliminary ATLAS results at 13~TeV~\cite{atlasICHEP}, 
illustrate the potential of this observable once higher statistics are available. The current Run 2 of the LHC at 13 TeV will allow these observables to be studied with  
much higher precision, and will extend the kinematic range that can be probed to larger values of the transverse momentum. 

\begin{figure}
  \centering
\includegraphics[width=7cm,natwidth=610,natheight=642]{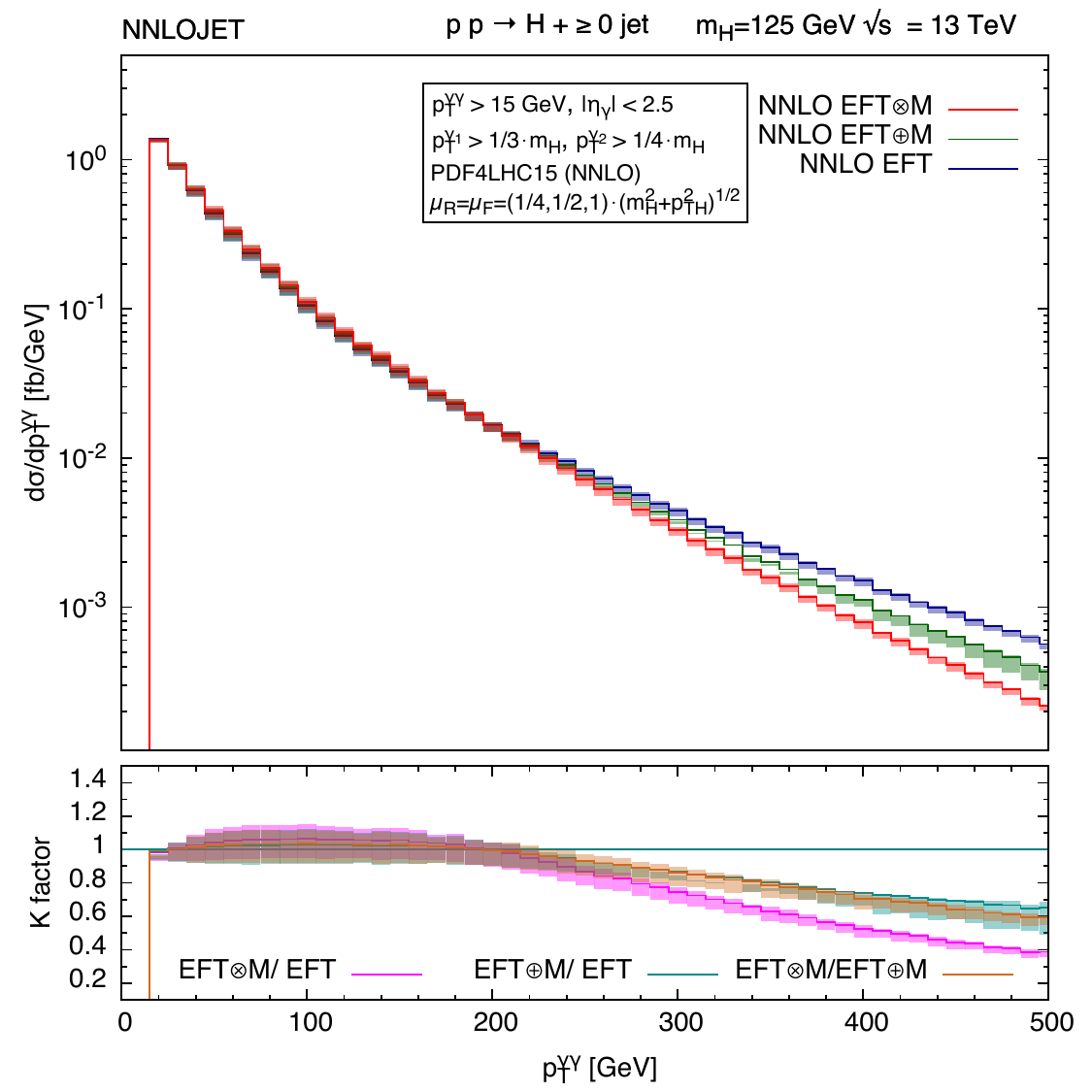}
\includegraphics[width=7cm,natwidth=610,natheight=642]{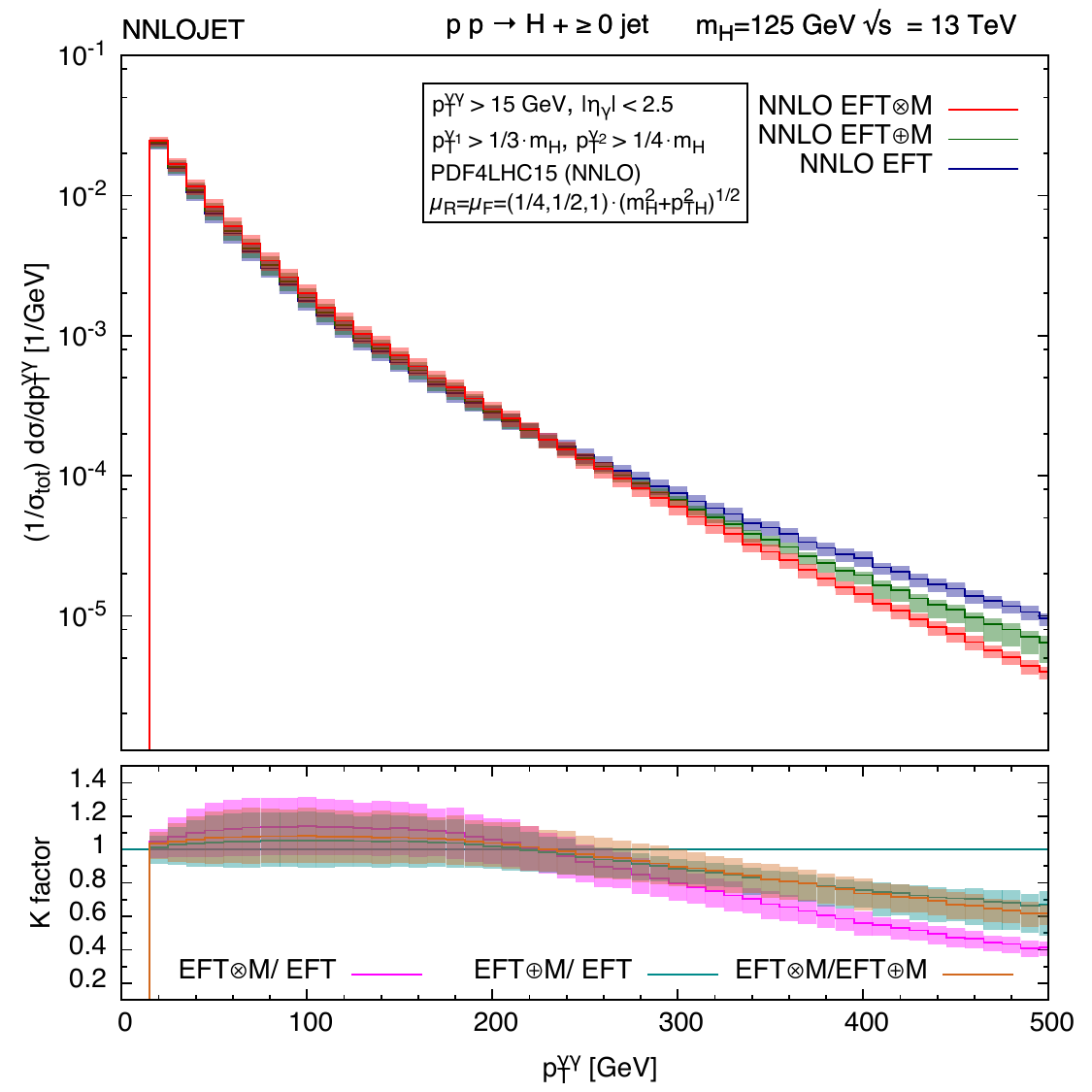}
\caption{Transverse momentum distribution of the Higgs boson at 13 TeV (for CMS fiducial cuts) 
for the EFT (red), \EFTplus\ (green) and \EFTtimes\ (blue) approximations. Left panel is the absolute cross sections, right panel normalized 
 to $\sigma_{H}$.
\label{fig:nnloweightpth}}
\end{figure}

To quantify the impact of the top quark mass effects,  we use
the CMS fiducial cuts and the theory parameters described in Section~\ref{sec:hjet} at 13 TeV. 
As discussed earlier, we consider two approximate approaches to estimating the mass effects defined in Eqs.~\eqref{eq:Mtimes} and \eqref{eq:Mplus}, the multiplicative \EFTtimes\ and additive \EFTplus\ approximations respectively in addition to the EFT in the large quark mass limit.
To quantify the uncertainty on these procedures, we compare in Figure~\ref{fig:nnloweightpth}
the \EFTplus\ (green) and  the \EFTtimes\ (red) predictions obtained according to Eqs.~\eqref{eq:Mplus} and \eqref{eq:Mtimes}. 
The EFT and \EFTtimes\ predictions (and the corresponding scale uncertainty) are simply related by $R(p_T)$ as shown in Fig.~\ref{fig:masslo8and13}(right). For Higgs transverse momentum $p_T^H > 200$~GeV, the EFT distribution is much harder than the \EFTtimes\ prediction, and as a result, the \EFTplus\ prediction lies between the two. 

The inclusion of quark mass effects at LO leads to a damping of the transverse momentum spectrum. Consequently, 
 in the \EFTplus\ prediction at large transverse momenta, the harder higher order EFT corrections
dominate over the  softer LO contribution with exact mass dependence. Even if the yet unknown NLO corrections to the exact mass 
dependence turn out to be numerically large, there is no reason for them to increase substantially with transverse momentum. The 
\EFTplus\ is therefore overestimating the hardness of the mass-corrected transverse momentum spectrum, and can 
 thus be considered as 
upper bound on the actual exact mass dependence. 
The  \EFTtimes\ prediction is on the other hand reweighting the full spectrum with the softness of 
the LO mass dependence of the $(H+1)$-parton process. A recent study~\cite{greiner} of the LO quark mass effects in Higgs-plus-multijet production 
demonstrated that the mass-dependent suppression (with respect to the EFT prediction) 
of large transverse momentum configurations is less strong for the $(H+2)$-parton and 
$(H+3)$-parton processes than it is for the $(H+1)$-parton process.
Consequently,  \EFTtimes\ could be considered as lower bound on the exact mass dependence.

 Lacking the full mass dependence of the predictions at NLO, it  is however premature to conclude 
on whether   \EFTplus\  or \EFTtimes\  should be considered to be more reliable. Instead, their spread serves to quantify the large 
systematic uncertainty that persists on the theoretical prediction of the transverse momentum distribution at high $p_T^H$. 
The difference between the different approaches increases with increasing $p_T^H$ and clearly exceeds the scale uncertainty for $p_T^H > 250$~GeV.  At $p_T^H\sim 400(500)$~GeV, the NNLO \EFTtimes\ approximation is $52\%$ ($39\%$) of the NNLO EFT with a small scale uncertainty.  Conversely, the \EFTplus\ has a much larger scale uncertainty and is roughly $74\%$ ($65\%$) of the NNLO EFT prediction. The  \EFTplus\  is larger than  \EFTtimes\ by a 
factor 1.42 (1.69), thereby estimating the uncertainty on the predictions in this large transverse momentum region. 

\begin{figure}
  \centering
\includegraphics[width=7cm,natwidth=610,natheight=642]{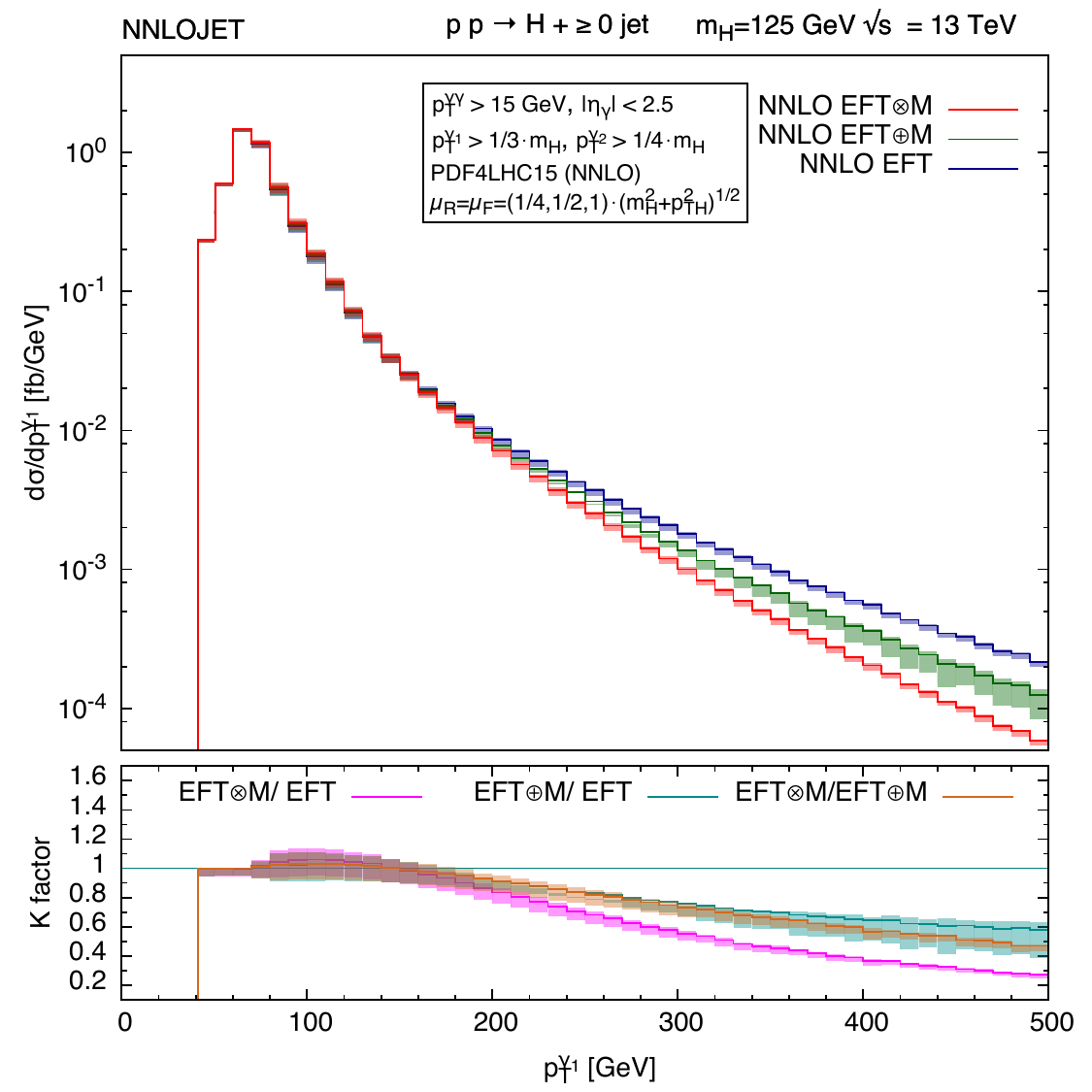}
\includegraphics[width=7cm,natwidth=610,natheight=642]{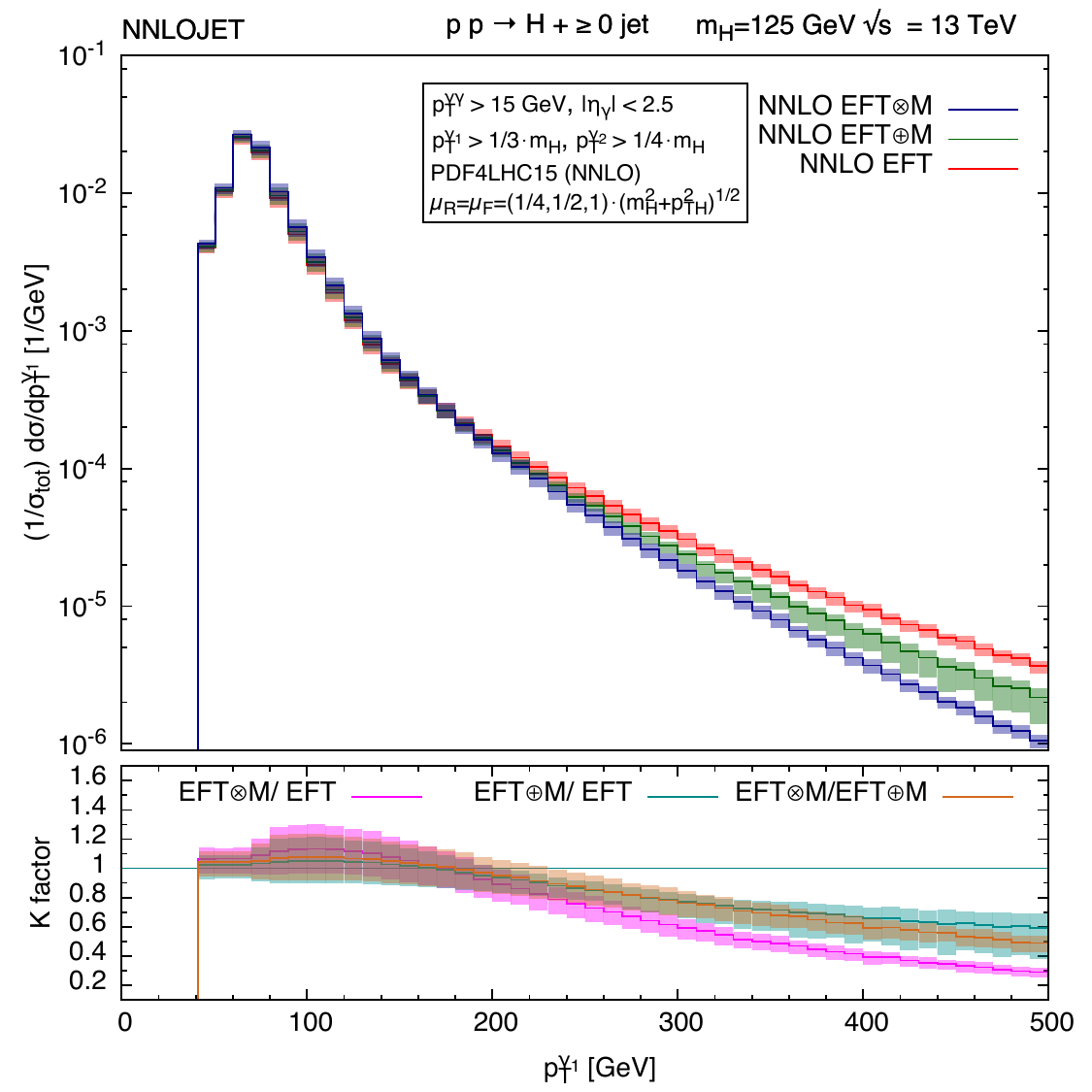}
\caption{Transverse momentum distribution of the leading photon at 13 TeV (for CMS fiducial cuts) for the EFT (red), \EFTplus\ (green) and \EFTtimes\ (blue) approximations. Left panel is the absolute cross sections, right panel normalized 
 to $\sigma_{H}$.
\label{fig:nnloweightptg1}}
\vspace{3mm}
 
\includegraphics[width=7cm,natwidth=610,natheight=642]{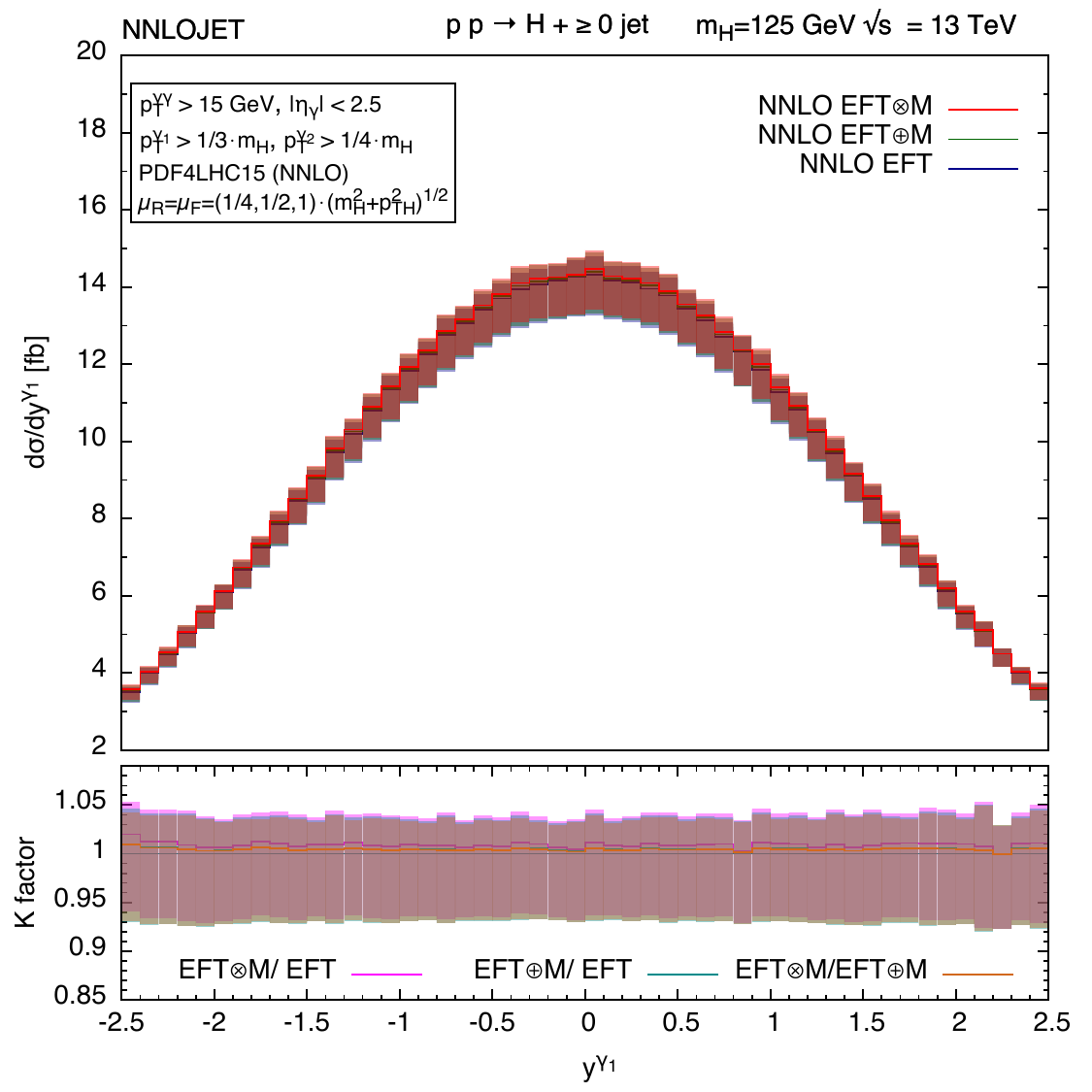}
\includegraphics[width=7cm,natwidth=610,natheight=642]{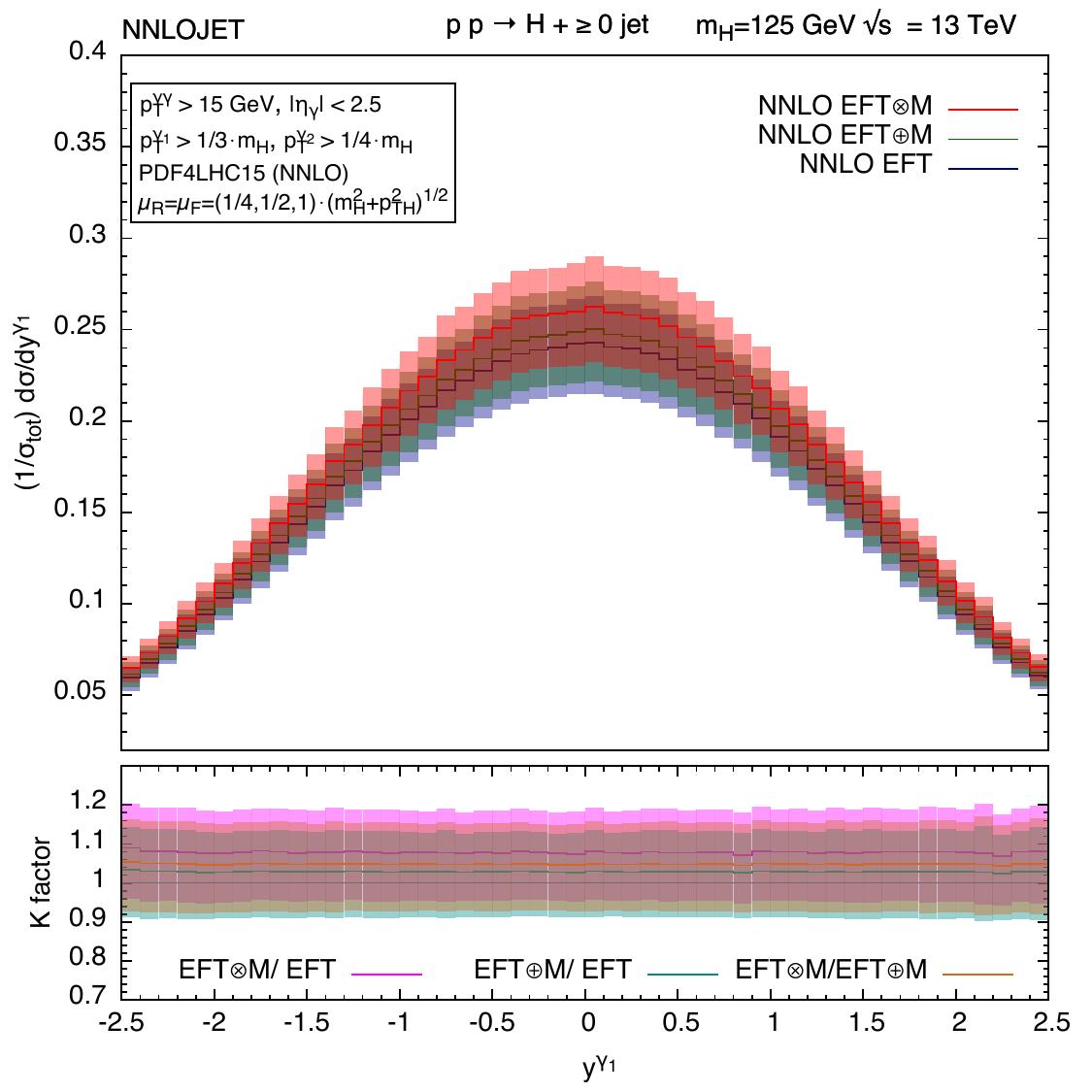}
\caption{Rapidity distribution of the leading photon at 13 TeV (for CMS fiducial cuts) for the EFT (red), \EFTplus\ (green) and \EFTtimes\ (blue) approximations. Left panel is the absolute cross sections, right panel normalized 
 to $\sigma_{H}$.
\label{fig:nnloweightyg1}}
\end{figure}

The behaviour of the Higgs boson transverse momentum 
distribution is mirrored in the transverse momentum distribution of the leading photon, shown in Fig.~\ref{fig:nnloweightptg1}. 
Again, the difference between the approximations are clearly visible. Above $p_T^{\gamma_1}\sim m_t$, NNLO \EFTtimes\ distribution is significantly softer than the NNLO EFT prediction, while the \EFTplus\ distribution lies between the two.  

The rapidity distribution of the leading photon, $y^{\gamma_1}$, is shown in Fig.~\ref{fig:nnloweightyg1}. Since this 
distribution is inclusive on the Higgs boson transverse momentum, it is dominated by its low transverse momentum region, starting at $p_T^H = 15$~GeV,
where the cross section is largest. In this region, the heavy quark loops are not resolved, and all three approximations yield very 
similar results, with similar scale uncertainties. The slight offset 
between the approximations for the normalized cross sections reflects the ratio between the inclusive $R$-factor at NNLO accuracy (from table ~\ref{tab:xsec}) and its differential value 
at the lower transverse momentum cut-off.

\section{Conclusions and outlook}
\label{sec:conc}

In this paper we have made a detailed study of Higgs production at large transverse momentum, both inclusively in the accompanying QCD radiation and in the presence of a jet. Our baseline calculation is at NNLO in the EFT approach, where the heavy quark loop that mediates the gluon fusion process 
is integrated out. The NNLO QCD corrections are found to be moderate and positive, they 
lead to a substantial reduction of the theory uncertainty on the predictions and open the way to precision studies with Higgs-plus-jet final states 
and on the Higgs boson transverse momentum distribution.

However, 
the kinematic regions probed by the LHC are influenced by the non-pointlike nature of the heavy quark loop that couples the Higgs boson to the gluons.  In this domain, it is vital that the effects of the top loop are captured as precisely as possible, and we therefore introduced two approximate descriptions that merge the NNLO corrections within the EFT together with the exact mass dependence that is known only at LO, the \EFTtimes\ and \EFTplus\ approximations. 
At small transverse momenta, the \EFTtimes\ and \EFTplus\ approximations are very similar and lead to a small enhancement 
of the predictions compared to the pure EFT.  
However, with increasing
 transverse momenta, the non-pointlike nature of the heavy quark loop becomes resolved and leads to a large suppression of the rate, 
 resulting in a softening of the transverse momentum spectrum as compared to the EFT prediction.  
   This suppression is more severe for the \EFTtimes\ approximation than for the \EFTplus\ approximation,
    and the difference between both should be considered as an  estimate of the current uncertainty at large $p_T^H$.

We made a detailed comparison of the available (statistics limited) ATLAS and CMS 8 TeV data for fiducial cross sections for Higgs production at moderate
 $p_T^H$ and in Higgs-plus-jet associated production.  In the kinematical region covered by these measurements, all three 
 approximations for the mass effects yield very similar results. The shape of the data for both experiments is well described at NNLO, 
 although the ATLAS data generally lies above the predictions.  The situation is improved by normalising the data to the inclusive Higgs cross section, although at the expense of increasing the theoretical scale uncertainty. Recent preliminary 13~TeV results from ATLAS are of a comparable 
 statistical quality to the 8~TeV data set, and agree well with the theory predictions both in shape and normalisation. 

To prepare for the larger data set expected from 
 Run 2 at 13 TeV, we made predictions for the Higgs transverse momentum distribution out to $p_T^H \sim 500$~GeV.  For observables 
 at 13~TeV that are dominated by low transverse momenta, top quark mass effects are moderate, and the different prescriptions agree very well, thus 
 providing reliable predictions.  
At large transverse momenta, the differences between the theory approximations for the top quark mass effects
 are large and lead to an uncertainty of ${\cal O}(50\%)$, which persists in any distribution that probes the kinematical region of large transverse momenta. A meaningful reduction of this uncertainty requires 
 knowledge of the full top quark mass dependence at NLO, which in turn demands the 
 two-loop  corrections to the Higgs-plus-three-parton amplitudes with massive internal quarks~\cite{Bonciani:2016qxi}.

\acknowledgments

We thank Thomas Morgan, Alexander Huss, Aude Gehrmann-De Ridder, 
Joao Pires, James Currie and Jan Niehues for useful discussions and their many contributions to the \NNLOJET\ code. We thank Fabrizio Caola and Markus Schulze for their assistance in comparing with the results of Ref.~\cite{caolaH2}. We thank the University of Zurich S3IT ({\tt http://www.s3it.uzh.ch}) for providing support and computational resources. XC thanks the IPPP at the University of Durham for hospitality and Li Lin Yang for useful discussions about the scale dependent terms. 
This research was supported in parts by the Swiss National Science Foundation (SNF) under contract 200020-162487, 
the UK Science and Technology Facilities Council, 
by a grant from the Swiss National Supercomputing Centre (CSCS) under project ID UZH10, by the Alexander von Humboldt Foundation in 
the framework of the Sofja Kovalevskaja Award 2014, and 
by the Research Executive Agency (REA) of the European Union under the Grant Agreement PITN-GA-2012-316704  (``HiggsTools'') 
and the ERC Advanced Grant MC@NNLO (340983).

\begin{appendix}
\section{Scale dependence of the cross section at NNLO in QCD}
\label{app:scales}

In the calculation of cross sections to fixed order in perturbation
theory, one has to fix the renormalization scale $\mu_R$
for the strong coupling constant $\alpha_s(\mu_R)$ as  well as for other effective couplings and mass parameters,
and the mass factorization scale $\mu_F$ for the parton 
distribution functions $f_i(x,\mu_F)$. 

The behaviour of the coupling constant and parton distributions 
under scale variations is determined by evolution equations, which resum scale-dependent logarithms to all orders in the coupling constant. 
The hard scattering cross sections are typically computed at a fixed pre-defined scale. 
Their scale-dependent terms can then be inferred by expanding the solutions of the evolution equations 
in powers of the strong coupling constant. In this appendix, we collect all formulae that are relevant to determine the scale-dependence of 
hadron collider cross sections to NNLO in QCD. In the EFT with a point-like Higgs boson coupling to gluons, the renormalisation scale dependence 
is modified, as explained in Section~\ref{sec:setup} above. 

For the strong coupling constant, the evolution equation reads:
\begin{eqnarray}
\mu_R^2 \frac{\d \alpha_s(\mu_R)}{\d \mu_R^2} &=& -\alpha_s(\mu_R) 
\left[\beta_0 \left(\frac{\alpha_s(\mu_R)}{2\pi}\right) 
+ \beta_1 \left(\frac{\alpha_s(\mu_R)}{2\pi}\right)^2 
+ \beta_2 \left(\frac{\alpha_s(\mu_R)}{2\pi}\right)^3 
+ {\cal O}(\alpha_s^4) \right]\,,\nonumber \\
\label{eq:running}
\end{eqnarray}
with the $\overline{{\rm MS}}$-scheme coefficients
\begin{eqnarray}
\label{eq:betas}
\beta_0 &=& \frac{11 \CA - 4 T_R \NF}{6}\;,\nonumber  \\
\beta_1 &=& \frac{17 \CA^2 - 10 C_A T_R \NF- 6C_F T_R \NF}{6}\;, \nonumber \\
\beta_2 &=&\frac{1}{432}
\big( 2857 C_A^3 + 108 C_F^2 T_R N_F -1230 C_FC_A T_R N_F
-2830 C_A^2T_RN_F \nonumber \\ &&
+ 264 C_FT_R^2 N_F^2 + 316 C_AT_R^2N_F^2\big)\;.
\end{eqnarray}

Using the solution of this 
equation, the coupling at a fixed scale $\mu_0$ can be expressed 
in terms of the coupling at $\mu_R$ by introducing 
\begin{equation}
L_R = \log \left(\frac{\mu_R^2}{\mu_0^2}\right)\,
\end{equation}
as
\begin{equation}
\alpha_S(\mu_0) = \alpha_s(\mu_R) \left[1 + \beta_0 L_R
\frac{\alpha_s(\mu_R)}{2\pi} + \left[\beta_0^2L^2_R + \beta_1L_R \right] 
\left(\frac{\alpha_s(\mu_R)}{2\pi}\right)^2 + {\cal O} (\alpha_s^3) 
\right] \,.
\label{eq:asfix}
\end{equation}

The perturbative expansion of a cross section involving $n+2$ partons at leading order
 starts at 
$\alpha_s^n$, provided that the 
Born process corresponds to tree level. The presence of the loop-level Born process, as is the case for Higgs production in 
gluon fusion modifies this counting, as discussed in Section~\ref{sec:setup} above. 

 In evaluating the expansion coefficients 
$\sigma^{(i)}=\sigma^{(i)}(\mu_0)$, 
the renormalization scale is fixed to a value $\mu_0$ (which can 
be dynamical event-by-event, rescalings can then be made 
for a fixed ratio $\mu_R/\mu_0$ for all events; e.g.\
if $\mu_0=p_{T,1}$, we can rescale to $\mu_R=2p_{T,1}$ or 
$\mu_R=p_{T,1}/2$, but not to $\mu_R=m_H$ or $\mu_R=H_T$). The expansion to NNLO reads:
\begin{equation}
\label{eq:sigmu0}
\sigma(\mu_0,\alpha_s(\mu_0)) 
= \left(\frac{\alpha_s(\mu_0)}{2\pi}\right)^n \sigma^{(0)}
+\left(\frac{\alpha_s(\mu_0)}{2\pi}\right)^{n+1} \sigma^{(1)}
+\left(\frac{\alpha_s(\mu_0)}{2\pi}\right)^{n+2} \sigma^{(2)} 
+{\cal O} (\alpha_s^{n+3}) \,.
\end{equation}
The scale dependence of the cross section can then be reconstructed 
by inserting \eqref{eq:asfix}:
\begin{eqnarray}
\label{eq:mRvar}
\lefteqn{\sigma(\mu_R,\alpha_s(\mu_R),L_R)}
\nonumber \\
&=& \left(\frac{\alpha_s(\mu_R)}{2\pi}\right)^n \sigma^{(0)}
+\left(\frac{\alpha_s(\mu_R)}{2\pi}\right)^{n+1} 
\left(\sigma^{(1)}+n\beta_0 L_R  \sigma^{(0)}\right)\nonumber \\ &&
+\left(\frac{\alpha_s(\mu_R)}{2\pi}\right)^{n+2} 
\left(\sigma^{(2)} + (n+1) \beta_0 L_R  \sigma^{(1)}+
n\beta_1 L_R \sigma^{(0)}+ \frac{n(n+1)}2 \beta_0^2 
L^2_R \sigma^{(0)}\right)\nonumber \\ &&
+{\cal O} (\alpha_s^{n+3}) \,.
\end{eqnarray}

The evolution of parton distributions is determined by the DGLAP evolution equation~\cite{ap}.
We omit the dependence on the Bjorken scaling variable $x$ here. 
\begin{equation}
\mu_F^2 \frac{\d}{\d \mu_F^2} f_i(\mu_F,\mu_R) 
= \sum_j P_{ij}(\alpha_s(\mu_R),\mu_F,\mu_R) \otimes f_j(\mu_F,\mu_R)\,,
\label{eq:apmaster}
\end{equation}
with the expansion to third order in terms of the 
splitting functions $P^{(n)}_{ij}$ computed at $\mu_F=\mu_R$
\begin{eqnarray}
P_{ij}(\alpha_s(\mu_R),\mu_F,\mu_R) & = &
\frac{\alpha_s(\mu_R)}{2\pi} P_{ij}^{(0)}  +
\left(\frac{\alpha_s(\mu_R)}{2\pi}\right)^2 
\left[ P_{ij}^{(1)} +\beta_0 l  P_{ij}^{(0)} 
\right]  \nonumber \\ && 
+
\left(\frac{\alpha_s(\mu_R)}{2\pi}\right)^3 
\left[ P_{ij}^{(2)} +\left(\beta_1   P_{ij}^{(0)} 
+ 2  \beta_0  P_{ij}^{(0)}\right) l + \beta_0^2 l^2  P_{ij}^{(0)}  
\right] + {\cal O} (\alpha_s^{4})\,,\nonumber \\
\label{eq:apsplit}
\end{eqnarray}
 where we introduced
\begin{equation}
l = \log \left(\frac{\mu_R^2}{\mu_F^2}\right).
\end{equation}

It is noteworthy that (\ref{eq:apsplit}) can be rewritten as
\begin{eqnarray}
P_{ij}(\alpha_s(\mu_R),\mu_F,\mu_R) & = &
\frac{\alpha_s(\mu_F)}{2\pi} P_{ij}^{(0)}  +
\left(\frac{\alpha_s(\mu_F)}{2\pi}\right)^2 
P_{ij}^{(1)} 
+
\left(\frac{\alpha_s(\mu_F)}{2\pi}\right)^3 
P_{ij}^{(2)} + {\cal O} (\alpha_s^{4})\,,\nonumber \\
\label{eq:apsplit1}
\end{eqnarray}
which implies that $f_i(\mu_F,\mu_R)$ and $f_i(\mu_F,\mu_F)$ fulfil the 
same evolution equation to all perturbative orders. The finite scheme 
transformation between both that one could postulate is thus vanishing 
to all orders, and 
both functions can at most vary in their non-perturbative boundary 
conditions (i.e.\ propagation of theory errors in fits of parton 
distribution functions). For all perturbative purposes, we thus have
\begin{equation}
f_i(\mu_F,\mu_R) = f_i(\mu_F,\mu_F) = f_i(\mu_F)\,,
\end{equation}
which we will normally use in all that follows (except if the 
scale transformation of the parton distribution is not expanded in 
$\alpha_s(\mu_F)$, but in $\alpha_s(\mu_R)$). 

The parton distribution at a fixed scale $\mu_0$ can be expressed in terms of parton 
distributions at $\mu_F$ by expanding the solution of (\ref{eq:apmaster}). We 
distinguish the expansion in powers of $\alpha_s(\mu_R)$ and in powers of $\alpha_s(\mu_F)$ and 
introduce
\begin{equation}
L_F = \log \left(\frac{\mu_F^2}{\mu_0^2}\right)\;. 
\end{equation}
The expansion in $\alpha_s(\mu_R)$ of the parton distribution at  $\mu_0$ reads then:
\begin{eqnarray}
f_i(\mu_0) & = & f_i(\mu_F) - \frac{\alpha_s(\mu_R)}{2\pi} P_{ij}^{(0)}\otimes
f_j(\mu_F) L_F \nonumber \\
&& - \left( \frac{\alpha_s(\mu_R)}{2\pi}\right)^2 \bigg[ P_{ij}^{(1)}\otimes
f_j(\mu_F) L_F - \frac{1}{2} P_{ij}^{(0)}\otimes P_{jk}^{(0)}\otimes f_k(\mu_F) L_F^2 
\nonumber \\ && \hspace{3cm}
+ P_{ij}^{(0)} \otimes f_j(\mu_F) \beta_0 L_F\left(l + \frac{1}{2} L_F \right)
\bigg] + {\cal O} (\alpha_s^{3})\;.
\label{eq:pdffix}
\end{eqnarray}
The expansion in powers of  $\alpha_s(\mu_F)$ is obtained from the above by 
setting $\mu_R=\mu_F$. 
\begin{eqnarray}
f_i(\mu_0) & = & f_i(\mu_F) - \frac{\alpha_s(\mu_F)}{2\pi} P_{ij}^{(0)}\otimes
f_j(\mu_F) L_F \nonumber \\
&& - \left( \frac{\alpha_s(\mu_F)}{2\pi}\right)^2 \bigg[ P_{ij}^{(1)}\otimes
f_j(\mu_F) L_F - \frac{1}{2} P_{ij}^{(0)}\otimes P_{jk}^{(0)}\otimes f_k(\mu_F) L_F^2 
\nonumber \\ && \hspace{3cm}
+\frac{1}{2} P_{ij}^{(0)} \otimes f_j(\mu_F) \beta_0 L_F^2\bigg] 
+ {\cal O} (\alpha_s^{3})\;.
\end{eqnarray}
In both expressions, a summation over indices appearing twice is implicit.

We compute the perturbative coefficients in a 
hadron collider cross section with default values of $\mu_F=\mu_R=\mu_0$.
The perturbative expansion to NNLO then reads:
\begin{eqnarray}
\sigma(\mu_0,\mu_0,\alpha_s(\mu_0)) 
&=& \left(\frac{\alpha_s(\mu_0)}{2\pi}\right)^n \hat{\sigma}_{ij}^{(0)}\otimes f_i(\mu_0) 
\otimes f_j(\mu_0) \nonumber \\ &&
+\left(\frac{\alpha_s(\mu_0)}{2\pi}\right)^{n+1} \hat{\sigma}_{ij}^{(1)}\otimes f_i(\mu_0) 
\otimes f_j(\mu_0)  \nonumber \\ &&
+\left(\frac{\alpha_s(\mu_0)}{2\pi}\right)^{n+2} \hat{\sigma}_{ij}^{(2)}\otimes f_i(\mu_0) 
\otimes f_j(\mu_0) 
+{\cal O} (\alpha_s^{n+3}) \,.
\end{eqnarray}

The full scale dependence of this expression can be recovered by inserting (\ref{eq:asfix}) and 
(\ref{eq:pdffix}) into the above:
\begin{eqnarray}
\lefteqn{\sigma(\mu_R,\mu_F,\alpha_s(\mu_R),L_R,L_F) 
=}\nonumber \\ &&
\left(\frac{\alpha_s(\mu_R)}{2\pi}\right)^n \hat{\sigma}_{ij}^{(0)}\otimes f_i(\mu_F) 
\otimes f_j(\mu_F) \nonumber \\ &&
+\left(\frac{\alpha_s(\mu_R)}{2\pi}\right)^{n+1} \hat{\sigma}_{ij}^{(1)}\otimes f_i(\mu_F) 
\otimes f_j(\mu_F)  \nonumber \\ &&
\hspace{5mm} +L_R \, \left(\frac{\alpha_s(\mu_R)}{2\pi}\right)^{n+1}  n\, \beta_0 \,
 \hat{\sigma}_{ij}^{(0)}\otimes f_i(\mu_F) 
\otimes f_j(\mu_F)  \nonumber \\ &&
\hspace{5mm} +L_F\, \left(\frac{\alpha_s(\mu_R)}{2\pi}\right)^{n+1}
\Big[-
 \hat{\sigma}_{ij}^{(0)}\otimes f_i(\mu_F) 
\otimes \left( P_{jk}^{(0)}\otimes f_k(\mu_F) \right)\nonumber \\ && \hspace{4cm}
-  \hat{\sigma}_{ij}^{(0)}\otimes  \left( P_{ik}^{(0)}\otimes f_k(\mu_F)\right) 
\otimes f_j(\mu_F) 
\Big] \nonumber \\ &&
+\left(\frac{\alpha_s(\mu_R)}{2\pi}\right)^{n+2} \hat{\sigma}_{ij}^{(2)}\otimes f_i(\mu_F) 
\otimes f_j(\mu_F) \nonumber \\ &&
\hspace{5mm} +L_R \, \left(\frac{\alpha_s(\mu_R)}{2\pi}\right)^{n+2}  \left((n+1)\, \beta_0 \,
 \hat{\sigma}_{ij}^{(1)} + n\, \beta_1 \, \hat{\sigma}_{ij}^{(0)}\right)
 \otimes f_i(\mu_F) 
\otimes f_j(\mu_F)  \nonumber \\ &&
\hspace{5mm} +L_R^2 \, \left(\frac{\alpha_s(\mu_R)}{2\pi}\right)^{n+2}  \frac{n(n+1)}{2}\, \beta^2_0 \,
 \hat{\sigma}_{ij}^{(0)}\otimes f_i(\mu_F) 
\otimes f_j(\mu_F)  \nonumber \\ &&
\hspace{5mm} +L_F\, \left(\frac{\alpha_s(\mu_R)}{2\pi}\right)^{n+2}
\Big[-
 \hat{\sigma}_{ij}^{(1)}\otimes f_i(\mu_F) 
\otimes \left( P_{jk}^{(0)}\otimes f_k(\mu_F) \right)\nonumber \\ && \hspace{4cm}
-  \hat{\sigma}_{ij}^{(1)}\otimes  \left( P_{ik}^{(0)}\otimes f_k(\mu_F)\right) 
\otimes f_j(\mu_F)   \nonumber \\ &&\hspace{4cm}
- \hat{\sigma}_{ij}^{(0)}\otimes f_i(\mu_F) 
\otimes \left( P_{jk}^{(1)}\otimes f_k(\mu_F) \right)\nonumber \\ && \hspace{4cm}
-  \hat{\sigma}_{ij}^{(0)}\otimes  \left( P_{ik}^{(1)}\otimes f_k(\mu_F)\right) 
\otimes f_j(\mu_F) 
\Big] \nonumber \\ &&
\hspace{5mm} +L_F^2\, \left(\frac{\alpha_s(\mu_R)}{2\pi}\right)^{n+2}
\Big[
 \hat{\sigma}_{ij}^{(0)}\otimes \left( P_{ik}^{(0)}\otimes f_k(\mu_F)\right) 
\otimes \left( P_{jl}^{(0)}\otimes f_l(\mu_F) \right)\nonumber \\ && \hspace{4cm}
+ \frac{1}{2} \hat{\sigma}_{ij}^{(0)}\otimes f_i(\mu_F)\otimes \left( 
P_{jk}^{(0)}\otimes P_{kl}^{(0)}\otimes f_l(\mu_F)\right)\nonumber \\ && \hspace{4cm}
+ \frac{1}{2} \hat{\sigma}_{ij}^{(0)}\otimes \left( 
P_{ik}^{(0)}\otimes P_{kl}^{(0)}\otimes f_l(\mu_F)\right)
\otimes f_j(\mu_F) \nonumber \\ && \hspace{4cm}
+\frac{1}{2} \beta_0 \,\hat{\sigma}_{ij}^{(0)}\otimes f_i(\mu_F) 
\otimes \left( P_{jk}^{(0)}\otimes f_k(\mu_F) \right)\nonumber \\ && \hspace{4cm}
+\frac{1}{2} \beta_0 \, \hat{\sigma}_{ij}^{(0)}\otimes  \left( P_{ik}^{(0)}\otimes f_k(\mu_F)\right) 
\otimes f_j(\mu_F)  \Big] \nonumber \\ &&
\hspace{5mm} +L_F L_R\, \left(\frac{\alpha_s(\mu_R)}{2\pi}\right)^{n+2}
\Big[- (n+1)\, \beta_0\,
 \hat{\sigma}_{ij}^{(0)}\otimes f_i(\mu_F) 
\otimes \left( P_{jk}^{(0)}\otimes f_k(\mu_F) \right)\nonumber \\ && \hspace{4cm}
- (n+1)\, \beta_0\, \hat{\sigma}_{ij}^{(0)}\otimes  \left( P_{ik}^{(0)}\otimes f_k(\mu_F)\right) 
\otimes f_j(\mu_F) \Big]  \nonumber \\ &&
+{\cal O} (\alpha_s^{n+3}) \,.
\end{eqnarray}

\end{appendix}

\end{document}